\documentclass[aps,prx,
reprint,
longbibliography]{revtex4-2}

\usepackage{amssymb,bm}
\usepackage{amsmath,eucal}\usepackage[normalem]{ulem}
\usepackage[usenames,dvipsnames]{color}
\usepackage{graphicx}
\usepackage{units}
\usepackage{dsfont}
\usepackage{hyperref}
\usepackage{enumitem}

\newcommand*{\VEC}[1]{\boldsymbol{#1}}

\newcommand*{\OP}[1]{{\cal{#1}}}
\newcommand*{\kB}{k_{\mathrm{B}}}

\renewcommand{\d}{\mathrm{d}}	

\newcommand*{\traj}[1]{\underline{#1}}
\newcommand*{\trajR}[1]{\underline{\tilde{#1}}}
\newcommand{\ti}{\tau_{\mathrm{i}}}
\newcommand{\tf}{\tau_{\mathrm{f}}}

\newcommand*{\Da}{D_{\mathrm{a}}}
\newcommand*{\ta}{\tau_{\mathrm{a}}}
\newcommand*{\pa}{p_{\mathrm{a}}}
\newcommand*{\po}{p_{\mathrm{0}}}
\newcommand*{\pint}{p_{\mathrm{int}}}

\newcommand*{\Pe}{\operatorname{Pe}}

\begin{document}

\title{Thermodynamic nature of irreversibility in active matter}

\author{Lennart Dabelow}
\affiliation{School of Mathematical Sciences, Queen Mary University of London, London E1 4NS, UK}

\author{Ralf Eichhorn}
\affiliation{Nordita, Royal Institute of Technology and Stockholm University,
106 91 Stockholm, Sweden}

\begin{abstract}
Active matter describes systems whose constituents convert energy from their surroundings into directed motion,
such as bacteria or catalytic colloids.
We establish
a thermodynamic law for dilute suspensions of
interacting active particles
in the form of a remarkable direct link between their dynamics and thermodynamics:
The rate at which irreversibility is built up in the non-equilibrium steady state
is a state function of the thermodynamic system parameters,
namely the number of particles, the temperature
and, as a distinctive characteristic of active matter,
the swim pressure.
Like in the famous fluctuation theorems of stochastic thermodynamics,
irreversibility is a dynamical measure that quantifies
the amount by which microscopic particle trajectories break time-reversal symmetry,
without reference to the typically unobservable processes underlying self-propulsion.
We derive the result for the paradigmatic model of active Ornstein-Uhlenbeck particles
with short-ranged repulsive interactions.
Based on numerical simulations and heuristic arguments,
we furthermore present a refined relation whose validity extends
from the dilute
into the phase-separated (MIPS) regime.
\end{abstract}
\date{\today}

\maketitle

\section{Introduction}
\label{sec:intro}

The fundamental goal of statistical mechanics is to understand the thermodynamics of macroscopic systems
based on the laws that govern their microscopic constituents.
A principal strategy is to connect
dynamical behavior to emergent thermodynamic properties.
Within the framework of stochastic thermodynamics~\cite{Jarzynski:2011eai, Seifert:2012stf, PelitiPigolotti:StochasticThermodynamics, Shiraishi:AnIntroductionToStochasticThermodynamics}, several such connections have been discovered for systems at micro- to nanometer scales,
which are strongly influenced by thermal fluctuations.
These results often go beyond the traditional realm of equilibrium systems and provide insights into systems \emph{far away from equilibrium}.
A seminal example is Seifert's detailed fluctuation theorem~\cite{Seifert:2005epa} for Brownian particles:
It relates the  thermodynamic entropy to the \emph{irreversibility} of the dynamics,
which
quantifies how strongly time-reversal symmetry is broken on the level of particle trajectories.
In its averaged form,
it generalizes the second law of thermodynamics,
which underlies essentially all applications in science and engineering that rely on thermodynamic processes.
Finding such connections between dynamics and thermodynamics is thus not only of fundamental interest
but also of immediate practical relevance.

Based on
such second-law-like notions,
irreversibility is often interpreted as a non-equilibrium generalization of entropy production \cite{Fodor:2022ibe},
even for systems for which its connection to thermodynamics is unclear.
A large and diverse class of such
non-equilibrium systems is so-called active matter
\cite{Romanczuk:2012abp,Bechinger:2016api,Takatori:2016:fsa, Marchetti:2016mmo,Martin:2021sma, Fodor:2022ibe}.
It includes, among others, many living organisms, catalytic colloids, and nanorobots.
The defining characteristic of active particles is their ability to self-propel by converting energy from their environment,
maintaining them in a permanent out-of-equilibrium state
even in the absence of external perturbations like fields and chemical or temperature gradients.
Moreover, they can exhibit a variety of intriguing phenomena such as clustering, swarming, and
motility-induced phase separation (MIPS)
\cite{Fily:2012aps, Cates:2013waa, Buttinoni:2013dcp, Palacci:2013lcl,Berthier:2013neg,Stenhammar:2013ctp,Buttinoni:2013dcp,Stenhammar:2014pba, Cates:2015mip, Takatori:2015tta,Szamel:2015gdo,Solon:2015ppe, Farage:2015eii, Winkler:2015vps, Bialke:2015nit, Flenner:2016tng,Levis:2017abe, Digregorio:2018fpd, Solon:2018gto, Caprini:2019cst,Martin:2021sma,Marchetti:2016mmo,Geyer:2019ffm, GrandPre:2021epf,Maggi:2021uco,Liao:2021evp, Kreienkamp:2022cfr, Li:2023tls}.
Relating the dynamics of active particles to their thermodynamic properties
is challenging because the dissipative processes generating self-propulsion,
and the entropy production associated with these processes, are
usually not directly accessible in experiment
and require additional, system-specific assumptions to be modeled in theory \cite{Pietzonka:2018epo, Markovich:2021taf, Fritz2023:tcm}.
However, the fact that the dynamics and observable macroscopic phenomenology
are very similar across systems with very different microscopic self-propulsion mechanisms
suggests that the particle trajectories alone encode relevant information about thermodynamic properties emerging on the macroscopic level of many
interacting active particles.
Nevertheless, and despite tenacious recent efforts
\cite{Fodor:2016hff, Mandal:2017epa, Puglisi:2017crf, Shankar:2018hep, Dabelow:2019iam, Caprini:2019epoa,Crosato:2019ies,Markovich:2021taf, Martin:2021sma,Datta:2022sla, Fodor:2022ibe},
it has remained largely elusive to find a thermodynamic interpretation of the
trajectory-wise irreversibility for active matter.

Here we discover such a remarkable thermodynamic law
for colloidal active particles.
Focusing on dilute suspensions of hard-core repulsive, self-propelled particles in two and three dimensions,
we first demonstrate analytically that the
average rate at which microscopic particle trajectories 
become increasingly irreversible while evolving in the non-equilibrium steady state
is a function of the thermodynamic system parameters
(Section \ref{sec:results:lowdensity}).
Besides the temperature and the number of particles,
these thermodynamic variables include a genuine characteristic of active matter, namely
the swim pressure or active pressure \cite{Takatori:2014sps, Yan:2015tsf,Solon:2015ppe,Winkler:2015vps, Bialke:2015nit, Speck:2016sta, Speck:2016ibp, Junot:2017avp, Levis:2017abe,Caprini:2018apc,Digregorio:2018fpd, Duzgun:2018abp,Speck:2020cfi},
i.e., the pressure contribution from the self-propulsion forces
(explicitly defined in Section \ref{sec:observables:pressure}).
The irreversibility is quantified by the log-ratio between the
probability of observing a specific time evolution of 
particle positions forward in time
and the probability of observing the same particle positions backward in time
(cf.~Section \ref{sec:observables:irrev}, in particular Eq.~\eqref{eq:Sigma}).
It
thus measures how strongly
microscopic
particle trajectories break time-reversal symmetry
\cite{Fodor:2016hff, Mandal:2017epa,Puglisi:2017crf,Shankar:2018hep,Dabelow:2019iam,Caprini:2019epoa, Flenner:2020amq,Markovich:2021taf, Martin:2021sma,Crosato:2019ies,Datta:2022sla,Fodor:2022ibe}.
Based on numerical simulations and heuristic
arguments,
we moreover obtain
a refined version of this thermodynamic
characterization of irreversibility
which extends beyond the dilute limit to
larger densities, if the activity is sufficiently high such that the particles undergo MIPS
(Section \ref{sec:results:MIPS}).
Our novel thermodynamic relation
lends itself to
widespread application in the analysis of
thermodynamic processes using
active matter as the working medium.

\begin{figure*}
\includegraphics[scale=1]{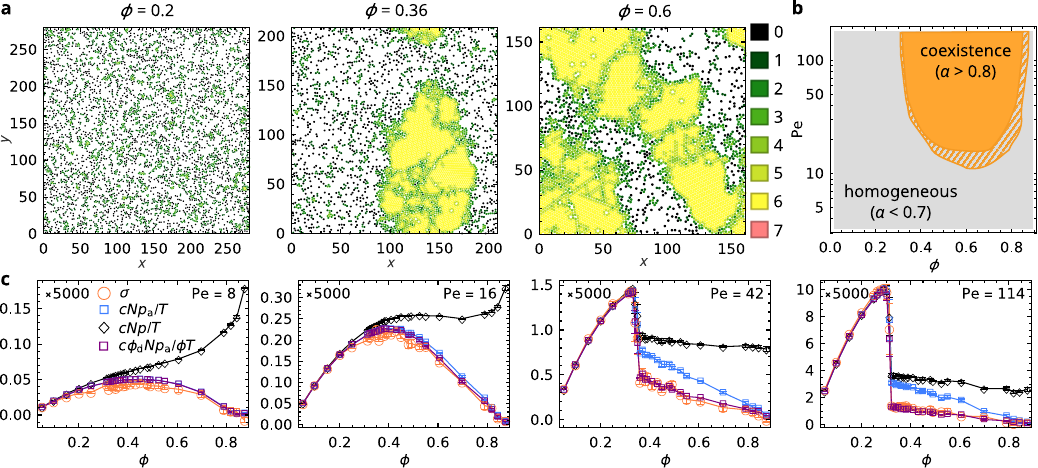}
\caption{Irreversibility and thermodynamic state variables
of $N = 5000$ active Orstein-Uhlenbeck particles
in $d=2$ dimensions.
(a) Snapshots of the dynamics
at P\'{e}clet number $\Pe = 42$ (defined above Eq.~\eqref{eq:OUP}) for packing fraction $\phi = 0.2$ (gas phase; defined above Eq.~\eqref{eq:LE}), $\phi = 0.36$ (critical point), and $\phi = 0.6$ (coexistence phase). Colors indicate the number of neighbors within a distance of $R/10$ with $R$ being the particle radius (see legend).
(b) 
Phase diagram in the $\phi$-$\Pe$ plane as determined by the giant number fluctuations exponent
$\alpha$ \cite{Fily:2012aps,Palacci:2013lcl}
(see Appendix~\ref{app:numerics:alpha} for details and raw data).
We classify situations with $\alpha < 0.7$ as homogeneous regimes and with $\alpha > 0.8$ as coexistence regimes;
in the shaded region, $0.7 \leq \alpha \leq 0.8$ and the available data are inconclusive.
(c) Directly computed irreversibility $\sigma$ (orange; defined in Eq.~\eqref{eq:sigma:def})
and its relation to the thermodynamic state variables:
$c N \pa / T$
for small $\phi$ (blue, cf.\ Eq.~\eqref{eq:sigma=pa});
$c N p / T$ for small $\phi$ and large $\Pe$ (black, cf.\ Eq.~\eqref{eq:sigma=p});
and $c (\phi_{\mathrm{d}}/\phi) N \pa / T$ in the MIPS regime (purple, cf.\ Eq.~\eqref{eq:sigma=pa:MIPS}).
The four panels show these quantities as a function of $\phi$
for various values of
$\Pe$;
$N$ is the number of particles, $T$ the temperature, $\pa$ the active pressure (defined in Eq.~\eqref{eq:pa}),
$p$ the total pressure (defined in Eq.~\eqref{eq:p}),
and $\phi_{\mathrm{d}}$ the packing fraction in the dilute phase (outside of the large clusters; defined in Eq.~\eqref{eq:phi:dilute}).
The prefactor $c = V_{\mathrm{p}} / (\kB \ta)$, obtained from Eq.~\eqref{eq:sigma=pa} for $d=2$, depends on the particles' size (volume) $V_{\mathrm p}$ and the persistence time $\ta$ of their self-propulsion (see Eq.~\eqref{eq:etaCorr}).
Data points are averages over steady-state trajectories of duration
$\tau = 2000$ and 6 independent repetitions of the simulations,
with the error bars indicating the corresponding standard error (see also Appendix~\ref{app:numerics}).
Simulation parameters:
$\mu = 0.1$ (dynamic viscosity), $\kB T = 0.01$, $\ta = 251$, interaction potential $U(r)$ from~\eqref{eq:IntPotentialNum} with $k = 0.2$, $R = 1$
(i.e., units of length can be determined relative to the particle radius $R=1$,
units of time relative to the self-diffusion time $R^2/D \approx 190$,
and units of force relative to $\kB T/R=0.01$);
$\phi$ and $\Pe$ are varied by changing the box size $L$ and the active diffusivity $\Da$, respectively.}
\label{fig:PhasesPressIrrev2D}
\end{figure*}

Figure~\ref{fig:PhasesPressIrrev2D} exemplarily summarizes our
findings for the two-dimensional case:
The dynamics and rich phenomenology of active matter are illustrated with
simulation snapshots (Panel a) and a phase diagram in the $\phi$-$\Pe$ plane (Panel b),
showing regimes of a homogeneous gas-like phase
and of a liquid-gas coexistence phase with MIPS.
The packing fraction $\phi$ is the fraction of the total volume occupied by the active particles.
The P\'{e}clet number $\Pe$ quantifies their activity.
Both quantities will be defined explicitly below (see above Eq.~\eqref{eq:LE} and above Eq.~\eqref{eq:OUP}).
Our main result is visualized in Panel~c:
The irreversibility production rate $\sigma$ (orange lines, defined in~\eqref{eq:sigma:def}) is
a function of the temperature $T$, the number of particles $N$ and the active pressure $\pa$ (blue lines, defined in~\eqref{eq:pa}) for small $\phi$ and all $\Pe$ ($\phi \lesssim 0.2$ in all
diagrams in Panel~c).
If the combination of thermodynamic state variables is rescaled by the ratio $\phi_{\mathrm{d}}/\phi$, where
$\phi_{\mathrm{d}}$ is the packing fraction of particles in the gas-like
(dilute) phase,
the connection between $\sigma$ and $N, \pa, T$ holds in the MIPS regime as well,
as can be seen by comparing the orange and purple lines for large $\Pe$ (third and fourth panels in Fig.~\ref{fig:PhasesPressIrrev2D}c).
This observation indicates that the irreversibility is dominated by particles in the dilute regions
with packing fraction $\phi_{\mathrm{d}}$, outside of the large clusters that form in the MIPS regime.
The physical reason is that particles within the bulk of these clusters 
become effectively immobilized, 
resulting in nearly indistinguishable forward and backward trajectories
that contribute negligibly to the irreversibility.
A more detailed discussion of this phenomenon is given in Section~\ref{sec:results:MIPS}.

\begin{figure*}
\includegraphics[width=\linewidth]{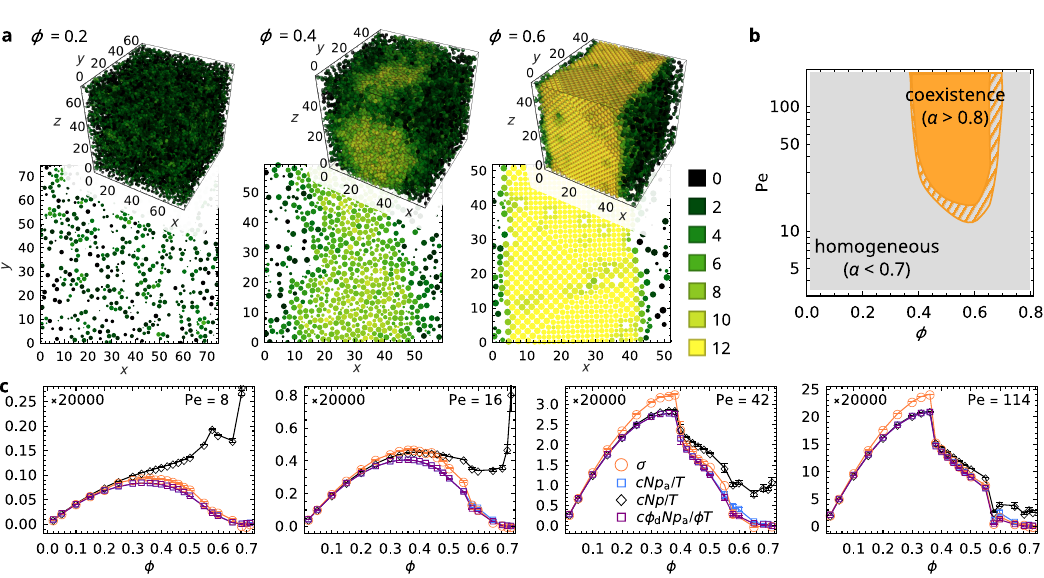}
\caption{Irreversibility and thermodynamic state variables
of $N = 20\,000$ active Orstein-Uhlenbeck particles
in $d = 3$ dimensions.
(a) Snapshots of the dynamics
at $\Pe = 294$ for $\phi = 0.2$, $\phi = 0.5$, and $\phi = 0.65$ (3D views and cuts through the $z = 24.5$ plane).
Colors indicate the number of neighbors within a distance of $R/10$ with $R$ being the particle radius
(see legend).
(b) 
Phase diagram in the $\phi$-$\Pe$ plane as determined by the giant number fluctuations exponent $\alpha$ \cite{Fily:2012aps,Palacci:2013lcl},
like in Fig.~\ref{fig:PhasesPressIrrev2D}.
(c) Directly computed irreversibility $\sigma$ (orange; defined in Eq.~\eqref{eq:sigma:def})
and its relation to the thermodynamic state variables:
$c N \pa / T$
for small $\phi$ (blue, cf.\ Eqs.~\eqref{eq:sigma=pa});
$c N p / T$ for small $\phi$ and large $\Pe$ (black, cf. Eq.~\eqref{eq:sigma=p});
and $c(\phi_{\mathrm{d}}/\phi) N \pa / T$ in the MIPS regime (purple, cf.\ Eq.~\eqref{eq:sigma=pa:MIPS}).
The four panels show these quantities as a function of
$\phi$ for various values of $\Pe$.
The prefactor $c = 2 V_{\mathrm{p}} / (\kB \ta)$, obtained from Eq.~\eqref{eq:sigma=pa} for $d=3$, depends on the particles' size (volume) $V_{\mathrm p}$ and the persistence time $\ta$ of their self-propulsion.
Other simulation details as in Fig.~\ref{fig:PhasesPressIrrev2D}
(see also Appendix~\ref{app:numerics}).
}
\label{fig:PhasesPressIrrev3D}
\end{figure*}

\section{Model}
\label{sec:model}

We consider a box of volume $V$ in $d = 2$ or $3$ dimensions
containing
$N$ interacting, spherical active particles
of radius $R$ in a solution at temperature $T$.
Hence the aforementioned packing fraction is $\phi := N V_{\mathrm{p}}/V$ with $V_{\mathrm{p}} = \pi R^2$ ($V_{\mathrm{p}} = 4 \pi R^3 / 3$) the particle volume in $d = 2$ ($d = 3$).

The active particles, labeled by the index $i=1,\ldots,N$, move according to the overdamped Langevin equation
\begin{equation}
\label{eq:LE}
\dot{\VEC{x}}_i(t) = \frac{1}{\gamma} \VEC{f}_i(\VEC{x}(t)) + \sqrt{2D} \, \VEC{\xi}_i(t) + \sqrt{2\Da} \, \VEC{\eta}_i(t) 
\, ,
\end{equation}
where $\VEC{x}_i(t)=(x_i^1(t),\ldots,x_i^d(t))$ is the position of particle $i$ at time $t$ (in $d$ dimensions),
and $\VEC{x}=\{\VEC{x}_i\}_{i=1}^N$ represents the set of all particle positions.
The dot denotes the derivative with respect to time $t$.
The total force $ \VEC{f}_i(\VEC{x}(t))$ acts on particle $i$ due to its interactions with all other particles.
We consider
strongly repulsive contact interactions between pairs of particles,
\begin{align}
\label{eq:force:def}
\VEC{f}_i(\VEC{x}) & = \sum_{j (\neq i)} \VEC{f}(\VEC{x}_i-\VEC{x}_j)
\, ,
\end{align}
which are derived from a
potential $U(r)$
such that $\VEC{f}(\VEC{r}) = -U'(\lvert \VEC{r} \rvert) \frac{\VEC{r}}{\lvert \VEC{r} \rvert}$,
with $U(r) \equiv 0$ if $r > 2R$
\cite{Stenhammar:2014pba,Solon:2015ppe, Solon:2015pin, Levis:2017abe, Digregorio:2018fpd, Martin:2021sma}.
Our main results presented in Sec.~\ref{sec:results} (and their analytical derivation) do
not depend on any further details of the potential, as long as it is sufficiently steep.

Thermal fluctuations in the aqueous solution
are modeled by mutually independent, unbiased Gaussian white
noise processes $\VEC{\xi}_i(t)=(\xi_i^1(t),\ldots,\xi_i^d(t))$
with $\langle \xi_i^\mu(t) \xi_j^\nu(t') \rangle = \delta_{ij} \delta^{\mu\nu} \delta(t-t')$.
Since the
solution represents a thermal environment at equilibrium, the thermal diffusion coefficient is given by the
Einstein relation $D=\kB T/\gamma$, with $\gamma$ being the viscous friction coefficient and $\kB$ Boltzmann's constant.

The self-propulsion of each active particle is modeled as a stochastic
velocity
$\sqrt{2\Da} \, \VEC{\eta}_i(t)$
whose typical magnitude is
set by the ``active diffusivity'' $\Da$, in analogy to the thermal diffusion coefficient $D$. This common approach to include “active forces” in the equation of motion \eqref{eq:LE}
(see, e.g.,
\cite{Romanczuk:2012abp, Speck:2016sta, Bechinger:2016api}
and references therein)
does not resolve the microscopic processes that drive self-propulsion, but rather provides an effective description
which reproduces the statistical properties of the active motion. The two key properties are:
(i) active fluctuations are persistent in time and space, and thus generate
non-Markovian dynamics, and
(ii) they break detailed balance and thus establish non-equilibrium conditions on the level of the individual active particles.
The P\'{e}clet number $\Pe := \sqrt{\Da / D}$  measures how strong the self-propulsion is compared to the thermal fluctuations.

A minimal model for such active fluctuations are
Ornstein-Uhlenbeck processes $\VEC{\eta}_i(t)=(\eta_i^1(t),\ldots,\eta_i^d(t))$.
They are generated from mutually independent,
unbiased, $\delta$-correlated Gaussian white noise sources $\VEC{\zeta}_i(t)=(\zeta_i^1(t),\ldots,\zeta_i^d(t))$
according to
\begin{equation}
\label{eq:OUP}
\dot{\VEC{\eta}}_i(t) = -\frac{1}{\ta} \VEC{\eta}_i(t) + \frac{1}{\ta} \VEC{\zeta}_i(t)
\, .
\end{equation}
These processes are exponentially correlated,
\begin{equation}
\label{eq:etaCorr}
\langle \eta_i^\mu(t) \eta_j^\nu(t') \rangle = \frac{\delta_{ij}\delta^{\mu\nu}}{2\ta} e^{-|t-t'|/\ta}
\, ,
\end{equation}
with correlation time $\ta$, which is thus a measure for the persistence of the active fluctuations.

The combination of \eqref{eq:LE} and \eqref{eq:OUP} represents so-called \emph{active Ornstein-Uhlenbeck particles}
(AOUPs) and serves as a standard model for studying
the (non-equilibrium) statistical mechanics of active matter 
\cite{Marconi:2015tas,Marconi:2017hta,Martin:2021sma,Fodor:2020dct,Flenner:2020amq,Caprini:2021fdr},
including fluctuation theorems
\cite{Shankar:2018hep,Dabelow:2019iam},
pressure
\cite{Marconi:2015tas, Sandford:2017pfe, Martin:2021sma},
and collective phenomena
\cite{Fodor:2016hff, Martin:2021sma, Caprini:2019cst, Maggi:2021uco,Li:2023tls,Saw:2023ctia}.
A commonly adopted simplification
is to neglect the thermal fluctuations
\cite{Szamel:2015gdo,Flenner:2016tng,Fodor:2016hff,Marconi:2017hta, Mandal:2017epa, Puglisi:2017crf,Sandford:2017pfe,Flenner:2020amq,Caprini:2021fdr,Martin:2021sma,Maggi:2021uco,Saw:2023ctia},
i.e., to set $D = 0$ in~\eqref{eq:LE}.
However, keeping both thermal and active fluctuations is crucial to
obtain
a thermodynamically consistent picture (see also Refs.~\cite{Puglisi:2017crf, Dabelow:2019iam}).

Compared to other models of active matter,
such as active Brownian particles (ABPs) or run-and-tumble particles (RTPs),
AOUPs exhibit similar large-scale phenomenology 
\cite{Fily:2012aps, Cates:2013waa, Farage:2015eii, Caprini:2019cst}.
For instance, they all exhibit regimes of motility-induced phase separation
(MIPS) at sufficiently high packing fractions and P\'{e}clet numbers
\cite{Fily:2012aps, Cates:2013waa,Buttinoni:2013dcp,Cates:2015mip, Takatori:2015tta, Marconi:2015tas, Solon:2015ppe, Farage:2015eii, Winkler:2015vps, Bialke:2015nit, Speck:2016sta, Fodor:2016hff, Levis:2017abe, Digregorio:2018fpd, Caprini:2019cst, Crosato:2019ies, Martin:2021sma, Solon:2018gto, Li:2023tls}.
In this phase, active particles gather in large clusters in some regions of the accessible volume,
even though there are no attractive interactions between them,
while other regions maintain a dilute gas-like appearance.
We illustrate this effect exemplarily in Figs.~\ref{fig:PhasesPressIrrev2D}a and~\ref{fig:PhasesPressIrrev3D}a.
Furthermore, despite their conceptual simplicity, AOUPs
proved suitable to describe various experiments
\cite{Maggi:2014gee, Argun:2016nbs}
and can be derived from a microscopic chemical-reaction model for self-propulsion \cite{Fritz2023:tcm}.

We therefore focus on the AOUP model \eqref{eq:LE}, \eqref{eq:OUP} in the following, expecting
(as argued in more detail in Sec.~\ref{sec:discussion} below)
that the physical essence of our results holds analogously
for other models of particulate active matter (e.g., ABPs and RTPs).

\section{Steady-state irreversibility and pressure}
\label{sec:observables}

We here introduce our two central observables and explain their physical relevance.

\subsection{Irreversibility of trajectories}
\label{sec:observables:irrev}

Inspired by
its well-known connection to entropy production for passive systems \cite{Seifert:2005epa},
we are interested in the thermodynamic
information encoded in the irreversibility of
particle trajectories,
while disregarding the unknown ``house-keeping'' entropy produced by the internal processes generating the self-propulsion. Hence, we
adopt the standard definition of the trajectory-wise irreversibility 
\cite{Fodor:2016hff, Mandal:2017epa,Puglisi:2017crf,Nardini:2017epf,Shankar:2018hep,Dabelow:2019iam,Caprini:2019epoa, GrandPre:2021epf,Markovich:2021taf, Martin:2021sma,Crosato:2019ies,Datta:2022sla,Fodor:2022ibe}
\begin{equation}
\label{eq:Sigma}
\Sigma[\traj{\VEC{x}}] := \ln \frac{P[\traj{\VEC{x}}]}{P[\trajR{\VEC{x}}]}
\, .
\end{equation}
Here, $P[\traj{\VEC{x}}]$ is the probability density to observe
the collection
of all $N$ particle trajectories $\traj{\VEC{x}}=\{ \VEC{x}(t) \}_{t=\tau_{\mathrm{i}}}^{\tau_{\mathrm{f}}}$ from an initial time
$\tau_{\mathrm{i}}$ to a final time $\tau_{\mathrm{f}}$ and,
correspondingly,
$P[\trajR{\VEC{x}}]$ is the probability density to observe the time-reversed
system ``history''
$\trajR{\VEC{x}}=\{ \VEC{x}(\tau_{\mathrm{i}}+\tau_{\mathrm{f}}-t) \}_{t=\tau_{\mathrm{i}}}^{\tau_{\mathrm{f}}}$, in which
all $N$ particles follow exactly the same trajectories as in the time-forward movement, but traced out
backward in time.

A positive (negative) value of $\Sigma[\traj{\VEC{x}}]$ indicates that the observed
forward-in-time history is
more (un)typical than its time-reversed counterpart, while 
$\Sigma[\traj{\VEC{x}}] = 0$ corresponds to seemingly time-reversible behavior.
Hence, the definition \eqref{eq:Sigma} provides a measure for the direction of time
observed on the level of particle histories,
motivating its designation as trajectory-wise \emph{irreversibility}.

The general idea behind studying $\Sigma[\traj{\VEC{x}}]$, Eq.~\eqref{eq:Sigma},
without reference to the self-propulsion processes $\VEC{\eta}(t)$,
is to inspect the thermodynamic information that is encoded in the particle trajectories
$\VEC{x}(t)$ alone.
This
approach complies with the usual experimental situation that only spatial positions are measurable
whereas details about interaction forces, self-propulsion drives and thermal fluctuations are inaccessible.
Furthermore,
the processes $\VEC{\eta}(t)$ are a phenomenological model for the particles' self-propulsion,
which reproduces, within certain limits, the statistical properties of active particle motion \cite{Maggi:2014gee, Argun:2016nbs}.
They are, however, not directly related to the chemical
and mechanical processes underlying self-propulsion,
so
even if an experimenter could measure $\VEC{\eta}(t)$,
the irreversibility derived from $\ln {P[\traj{\VEC{x}},\traj{\VEC{\eta}}]}/{P[\trajR{\VEC{x}},\trajR{\VEC{\eta}}]}$
would not capture the true dissipation or entropy production \cite{Fritz2023:tcm}.
Despite their vastly different propulsions mechanisms, various interacting
active particle systems exhibit similar large-scale behavior and phenomenology
in their trajectories $\VEC{x}(t)$ (see also below Eq.~\eqref{eq:etaCorr}),
suggesting that essential
aspects of the thermodynamics and statistical mechanics of active matter can be understood
from the particle trajectories alone,
without the need to know or observe the precise details of the self-propulsion mechanism.

Our first main quantity of interest is the \emph{average irreversibility produced per unit time} in the steady state of the system,
\begin{equation}
\label{eq:sigma:def}
\sigma := \lim_{\tau \to \infty} \frac{\left\langle \Sigma[\traj{\VEC{x}}] \right\rangle}{\tau}
\, ,
\end{equation}
where the
average $\langle \,\cdots \rangle$ is over all system trajectories,
and where we set $-\tau_{\mathrm{i}}=\tau/2=\tau_{\mathrm{f}}$ for convenience.
To simplify terminology,
we refer to $\sigma$ loosely as ``irreversibility''
or ``irreversibility production'' in the following.

Extending the single-particle expressions for the irreversibility from
\cite{Dabelow:2019iam, Caprini:2019epoa, Dabelow:2021iam}
to many identical, interacting particles is straightforward. We find
\begin{multline}
\label{eq:sigma:1}
\sigma = \lim_{\tau \to \infty}\frac{1}{\tau D}
\int_{-\tau/2}^{\tau/2} \!\! \d t \int_{-\tau/2}^{\tau/2} \!\! \d t^\prime \;
	\frac{1}{\gamma} \sum_{i=1}^N \left\langle \dot{\VEC{x}}_i(t) \cdot \VEC{f}_i(\bm x(t^\prime)) \right\rangle
\\ \mbox{} \times
	\left[ \delta(t-t^\prime) - \frac{\Da}{D} \Gamma(t, t^\prime) \right]
\, , 
\end{multline}
where the non-local integration kernel $\Gamma(t, t^\prime)$ is given
in Appendix~\ref{app:sigma}, Eq.~\eqref{eq:Gamma}.
This non-local kernel and the resulting
double-integral structure of Eq.~\eqref{eq:sigma:1} are a consequence of
the non-Markovian nature of the process $\VEC{x}(t)$ due to the finite correlation time of the active fluctuations.
As explained in Appendix~\ref{app:sigma},
we can perform the $\tau \to \infty$ limit in \eqref{eq:sigma:1},
leading to
\begin{multline}
\label{eq:sigma:2}
\sigma =  \frac{\Da}{2 D^2 \ta^2}
	\int_0^\infty \!\! \mathrm{d}s \, e^{-s/\tau_*}
	\left[
		\sum_{i=1}^N \frac{1}{\gamma} \left\langle \VEC{x}_i(t_0) \cdot \VEC{f}_i(\VEC{x}(t_0+s)) \right\rangle
	\right.
\\
	\left. \mbox{}
	      -	\sum_{i=1}^N \frac{1}{\gamma}  \left\langle \VEC{x}_i(t_0) \cdot \VEC{f}_i(\VEC{x}(t_0-s)) \right\rangle
	\right]
\, ,
\end{multline}
where $t_0$ is an arbitrary reference time, and $\tau_* := \ta/\sqrt{1+\Da/D}$.
Note that throughout this article all products between stochastic processes are tacitly
understood to be interpreted in the Stratonovich sense.

\emph{A priori}, there is no
obvious
connection between the irreversibility $\sigma$ from~\eqref{eq:sigma:2} and the thermodynamical properties of the system.
The main aim of this article is to establish such a connection
for the non-equilibrium steady state of
active-particle suspensions.

\subsection{Pressure}
\label{sec:observables:pressure}

A second key
observable is the total pressure $p$ exerted by the active particles on the container walls. It
can be decomposed into three contributions 
\cite{Takatori:2014sps, Solon:2015ppe, Marconi:2015tas, Solon:2015pin, Winkler:2015vps, Bialke:2015nit, Nikola:2016apw, Speck:2016sta, Junot:2017avp, Levis:2017abe, Sandford:2017pfe,Marconi:2017hta, Digregorio:2018fpd, Speck:2020cfi, Martin:2021sma,Speck:2016ibp},
\begin{equation}
\label{eq:p}
p = \po + \pint + \pa
\, ;
\end{equation}
note that we disregard the pressure coming from the aqueous solution as an irrelevant reference pressure.
It is known that, for our current setup, the pressure $p$ is a state function
\cite{Takatori:2014sps, Solon:2015ppe, Solon:2015pin, Winkler:2015vps, Digregorio:2018fpd, Nikola:2016apw}.
The various contributions in \eqref{eq:p} can then be
calculated from the virials of the forces appearing in~\eqref{eq:LE}
\cite{Takatori:2014sps, Winkler:2015vps, Speck:2016sta, Levis:2017abe, Digregorio:2018fpd,Speck:2016ibp}.
The standard osmotic pressure or ideal-gas contribution from the thermal fluctuations is
\begin{subequations}
\label{eq:pressures}
\begin{equation}
\label{eq:p0}
p_0 = \frac{\gamma}{dV} \sum_{i=1}^N \langle \VEC{x}_i(t_0) \cdot \sqrt{2D}\,\VEC{\xi}_i(t_0) \rangle
	= \frac{N \kB T}{V}
\, .
\end{equation}
The interaction pressure $\pint$
is the contribution stemming from the particle-particle interactions,
\begin{equation}
\label{eq:pint}
	\pint = \frac{1}{d V} \sum_{i=1}^N \langle \VEC{x}_i(t_0)  \cdot \VEC{f}_i(\VEC{x}(t_0)) \rangle \,.
\end{equation}
Finally,
the \emph{active pressure}, or swim pressure, is
the component generated by the self-propulsion forces
\cite{Takatori:2014sps, Yan:2015tsf,Solon:2015ppe,Winkler:2015vps, Bialke:2015nit, Speck:2016sta, Speck:2016ibp, Junot:2017avp, Levis:2017abe,Digregorio:2018fpd, Speck:2020cfi},
\begin{equation}
\label{eq:pa}
	\pa = \frac{\gamma}{d V} \sum_{i=1}^N \langle \VEC{x}_i(t_0) \cdot \sqrt{2 \Da} \VEC{\eta}_i(t_0) \rangle
\, .
\end{equation}
\end{subequations}

With the relations \eqref{eq:pressures}, the total pressure is given as a \emph{mechanical pressure}
in terms of the micromechanical forces in the
equations of motion \eqref{eq:LE}, and is thus valid even out of equilibrium \cite{Takatori:2014sps}.
For this reason it is a key concept in the study of active matter 
\cite{Takatori:2014sps, Solon:2015ppe, Marconi:2015tas, Solon:2015pin, Winkler:2015vps, Bialke:2015nit, Nikola:2016apw, Speck:2016sta, Junot:2017avp, Levis:2017abe, Sandford:2017pfe, Marconi:2017hta,Digregorio:2018fpd, Speck:2020cfi, Martin:2021sma,Speck:2016ibp,Cameron:2023eos},
and serves as a powerful tool towards understanding the
thermodynamic(-like) properties of active particle systems
\cite{Takatori:2014sps,Solon:2018gto,Levis:2017abe,Junot:2017avp,Takatori:2015tta, Marconi:2015tas,Cameron:2023eos}.

\begin{figure}
\includegraphics[scale=1]{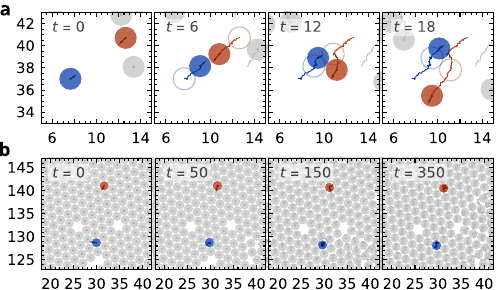}
\caption{Examples of collision events and their (ir)reversibility. The frames show time-series snapshots from numerical simulations
of the setup in
Fig.~\ref{fig:PhasesPressIrrev2D} (steady state, $\phi = 0.6$, $\Pe = 42$),
highlighting the trajectories of two specific particles (red and blue).
Empty circles
indicate their positions in the previous snapshot
and arrows the direction and magnitude of their self-propulsion.
(a) Pair collision:
Two particles approach each other, recoil and shift during the collision,
and finally
pass each other, maintaining roughly the same direction of motion as before the collision.
The time asymmetry becomes apparent by comparing to the hypothetical backward process,
where the two particles would first pass each other,
but then suddenly shift and align before continuing in their
initial direction of motion again.
(b) Bulk of a cluster:
Particles collide frequently with their neighbors, but barely move,
leaving no significant traces of irreversibility in their trajectories.
}
\label{fig:CollisionTrajectories}
\end{figure}

\section{Thermodynamics of Irreversibility}
\label{sec:results}

In interacting active matter systems, the two key parameters characterizing the transition to
MIPS are the packing fraction $\phi$ and the P\'{e}clet number $\Pe$
(see above Eq.~\eqref{eq:LE} and above Eq.~\eqref{eq:OUP}).
Hence, we analyze the (thermodynamic) properties of the irreversibility as a function of these two parameter.

\subsection{Low density}
\label{sec:results:lowdensity}

We focus on the limit of small packing fractions $\phi$ first,
representing the paradigmatic situation of a dilute ``active particle gas'' \cite{Shankar:2018hep}.
In this limit,
the integrals over the correlation functions involving two time points,
Eq.~\eqref{eq:sigma:2}, can be expressed as
equal-time correlation functions,
which can be shown in two steps.

First, we rewrite the correlation functions in~\eqref{eq:sigma:2} as
\begin{multline}
\label{eq:rel1}
\frac{1}{\gamma}
\sum_{i=1}^N
\langle \VEC{x}_i(t_0) \cdot \VEC{f}_i(\VEC{x}(t_0 \pm s)) \rangle
\\
= -\frac{2^{d-2}}{d} \phi
\sum_{i=1}^N
\langle 
	\VEC{x}_i(t_0)
	\cdot [\sqrt{2D} \, \VEC{\xi}_i(t_0 \pm s) + \sqrt{2\Da}  \,\VEC{\eta}_i(t_0 \pm s) ] \rangle
\\[-1ex]  
	\times \left( 1 + \OP{O}(\phi) \right)
\, .
\end{multline}
To derive this relation, we
exploit that, according to Eq.~\eqref{eq:sigma:2},
irreversibility is produced only during particle collisions, i.e., when
$\lvert \VEC x_i - \VEC x_j \rvert \leq 2R$ for some $i$ and $j$ ($i \neq j$),
because $\VEC{f}_i(\VEC{x}) = 0$ otherwise.
For small packing fractions $\phi$,
irreversibility production is dominated by pair-collision events
(cf.\ Fig.~\ref{fig:CollisionTrajectories}a),
because simultaneous collisions
of more than two particles are extremely rare.
Focusing on such pair collisions,
we replace
$\VEC f_i(\VEC{x})$ in $\langle \VEC{x}_i(t_0) \cdot \VEC{f}_i(\VEC{x}(t_0 \pm s)) \rangle$
by the sum of friction and fluctuating forces according to
the Langevin equation of motion \eqref{eq:LE}.
Due to the short-ranged, hard-core character of the
interactions, the contribution of the friction forces $-\gamma\dot{\VEC{x}}_i$ 
turns out to be negligibly small
during collision events,
such that only the fluctuating forces remain.
Likewise, the average $\langle \,\cdots \rangle$ over all possible positions of all particles
is effectively restricted to pair configurations with $|\VEC{x}_i-\VEC{x}_j| \leq 2R$,
leading to the appearance of the prefactor proportional to $\phi$ in~\eqref{eq:rel1}.
Details of the calculation are provided in Appendix~\ref{app:rel1}.

Second,
exploiting the replacement~\eqref{eq:rel1} in Eq.~\eqref{eq:sigma:2},
we evaluate the resulting integral over the correlations between particle positions and fluctuating forces,
\begin{multline}
\label{eq:rel2}
\int_0^\infty \d s \, \langle \VEC{x}_i(t_0)
	\cdot [ \sqrt{2D} \, \VEC{\xi}_i(t_0 \pm s) + \sqrt{2\Da}  \,\VEC{\eta}_i(t_0 \pm s) ] \rangle  \, 
	e^{-s/\tau_*}
\\
= \frac{\ta}{\sqrt{1+ \Da/D} \pm 1}
	\langle \VEC{x}_i(t_0) \cdot \sqrt{2\Da}\VEC{\eta}_i(t_0) \rangle
\, .
\end{multline}
This relation can be proven exactly with the help of Novikov's theorem \cite{Novikov:1965frf}
as detailed in Appendix~\ref{app:rel2}.

Using Eqs.~\eqref{eq:rel1} and~\eqref{eq:rel2}, the expression \eqref{eq:sigma:2} for the irreversibility turns into
\begin{equation}
\label{eq:sigma:state}
\sigma = \frac{2^{d-2}}{d\ta} \phi
\frac{\sum_{i=1}^N \langle \VEC{x}_i(t_0) \cdot \sqrt{2\Da}\VEC{\eta}_i(t_0) \rangle}{D}
\left(1 + \OP{O}(\phi) \right)
\, .
\end{equation}
With the relation~\eqref{eq:pa} for the active pressure,
we then arrive at our first main result,
\begin{equation}
\label{eq:sigma=pa}
	\sigma = \left( \frac{2^{d-2} V_{\mathrm{p}}}{\kB \ta} \right) \frac{ N \pa }{ T} \, [1 + \OP{O}(\phi)]
\, ,
\end{equation}
for spatial dimensions $d=2,3$.
The steady-state irreversibility production rate $\sigma$ is a function of the macroscopic thermodynamic
system parameters, namely the traditional state variables $N$ and $T$ as well as the active pressure $\pa$.
In addition, the particle volume $V_{\mathrm{p}}$ and the correlation time $\ta$ appear as measures of the
particles' interaction range and their characteristic reorientation time, respectively.

If
the activity is high ($\Pe \gg 1$) \emph{and} the packing fraction is small ($\phi \ll 1$),
we can furthermore relate the irreversibility to the \emph{total pressure} $p$.
In this limit,
$\pa$ dominates the osmotic pressure because $\frac{\pa}{\po} = \mathcal{O}(\Pe^2)$
and also the interaction pressure because $\frac{\pa}{\pint} = \mathcal{O}(\phi^{-1})$,
see Appendix~\ref{app:pa} for more details.
Consequently, we can approximate $\pa = p [1 + \OP{O}(\phi, \Pe^{-2})]$ and thus
\begin{equation}
\label{eq:sigma=p}
\sigma = \left( \frac{2^{d-2} V_{\mathrm{p}}}{\kB \ta} \right) \frac{ N p }{ T} \, [1 + \OP{O}(\phi, \Pe^{-2})]
\, ,
\end{equation}
which is our second main result.

\subsection{MIPS regime}
\label{sec:results:MIPS}

The relation~\eqref{eq:sigma=pa} is valid for small packing fractions
since Eq.~\eqref{eq:rel1} was derived in the limit of low particle density (cf.\ Appendix~\ref{app:rel1});
as indicated we estimate the relative error to be
${\mathcal{O}}(\phi)$.
For larger values of $\phi$,
the key assumption that most collisions are pair collisions breaks down
because simultaneous encounters
of three or more particles become significantly more likely,
leading to higher-order corrections in \eqref{eq:sigma=pa} in principle.
However,
even
though multi-particle collisions occur more frequently when $\phi$ becomes larger,
it
turns out that
pair collisions continue to dominate irreversibility production if the system exhibits motility-induced phase separation (MIPS). 
In this regime, a significant number of particles find themselves in large solid-like clusters  (see also Figs.~\ref{fig:PhasesPressIrrev2D} and~\ref{fig:PhasesPressIrrev3D}).
We denote the average number of particles in such
clusters by $N_{\mathrm{c}}$ and their average fraction by $n_{\mathrm{c}} := N_{\mathrm{c}} / N$.
The remaining $N - N_{\mathrm{c}}$ particles still move in a dilute, gas-like phase.

Particles in the bulk of the emerging
clusters
are
basically
immobilized
($\dot{\VEC x}_i(t) \approx 0$)
by the frequent multi-particle collisions with their neighbors
(cf.\ Fig.~\ref{fig:CollisionTrajectories}b).
Moreover, even though these particles collide very frequently,
the total force $\bm f_i$ on any single particle typically averages to zero even on short time scales.
It follows from~\eqref{eq:sigma:1}
that particles in the bulk of large clusters do not contribute significantly
to the irreversibility production.
This can also be understood intuitively:
Looking at a movie of particles that are essentially at rest,
it is impossible to tell whether such a recording is being played forward or backward.
Moreover, contributions from particle collisions with the surface
of the cluster are negligible compared to bulk effects
for sufficiently large systems.
Thus, irreversibility is predominantly produced within the dilute regions,
where pair collisions continue to prevail.
In other words, we can restrict the evaluation of the correlation function~\eqref{eq:rel1} to the dilute region of the phase-separated active particle system.
By the same arguments as in Appendix~\ref{app:rel1},
we then recover Eq.~\eqref{eq:rel1} provided that the total packing fraction $\phi$ is replaced by the packing fraction $\phi_{\mathrm{d}}$ of the dilute region.
Consequently, replacing $\phi_{\mathrm{d}}$ with $\phi$ also in Eq.~\eqref{eq:sigma:state},
we obtain as our third main result a corrected relationship between $\sigma$ and the thermodynamic state variables in the MIPS regime,
\begin{subequations}
\label{eq:sigma=pa:MIPS:both}
\begin{align}
\label{eq:sigma=pa:MIPS}
\sigma &\simeq  \left( \frac{2^{d-2} V_{\mathrm{p}}}{\kB \ta} \right) \left( \frac{\phi_{\mathrm{d}}}{\phi} \right) \frac{ N \pa }{ T} \\
\label{eq:sigma=pa:MIPS2}
	&\simeq \left( \frac{2^{d-2} V_{\mathrm{p}}}{\kB \ta} \right) \left( \frac{1 - n_{\mathrm{c}}}{1 - n_{\mathrm{c}} \phi / \phi_{\mathrm{cp}}} \right) \frac{ N \pa }{ T} \,.
\end{align}
\end{subequations}
Since $\phi_{\mathrm{d}} = \phi$ at low density in the absence of MIPS,
the relation~\eqref{eq:sigma=pa:MIPS} reduces to the low-density result~\eqref{eq:sigma=pa}
in the limit $\phi \ll 1$.
Hence Eq.~\eqref{eq:sigma=pa:MIPS} generalizes Eq.~\eqref{eq:sigma=pa} to a larger region of the $\phi$-$\Pe$ parameter space,
but the error in~\eqref{eq:sigma=pa:MIPS} is less well controlled.
Moreover, the relation~\eqref{eq:sigma=pa:MIPS} is expected to work better for larger systems
as it relies on
the aforementioned assumption that
effects on the surface of particle clusters
are negligible.

The alternative form~\eqref{eq:sigma=pa:MIPS2} of the right-hand side results from the following
estimate of
$\phi_{\mathrm{d}}$:
Since the $N_{\mathrm{c}}$ particles in the cluster
are essentially densely packed  (up to small defects),
the volume occupied by the cluster is
\begin{equation}
	V_{\mathrm{c}} \simeq \frac{N_{\mathrm{c}} V_{\mathrm{p}}}{\phi_{\mathrm{cp}}} = \frac{n_{\mathrm{c}} \phi}{\phi_{\mathrm{cp}}} V \,,
\end{equation}
where $\phi_{\mathrm{cp}}$ is the packing fraction for the closest packing of identical spheres
[$\phi_{\mathrm{cp}} = \pi / (2 \sqrt{3}) \approx 0.907$ in $d = 2$ and $\phi_{\mathrm{cp}} = \pi / (3 \sqrt{2}) \approx 0.74$ in $d = 3$ \cite{Hales:2000ch}].
The packing fraction $\phi_{\mathrm{d}}$ in the dilute phase is thus
\begin{equation}
\label{eq:phi:dilute}
	\phi_{\mathrm{d}} = \frac{(N - N_{\mathrm{c}}) V_{\mathrm{p}}}{V - V_{\mathrm{c}}}
	\simeq \phi \frac{1 - n_{\mathrm{c}}}{1 - n_{\mathrm{c}} \phi / \phi_{\mathrm{cp}}} \,.
\end{equation}
We note that $n_{\mathrm{c}}$, or, equivalently, $\phi_{\mathrm{d}}$, are not control parameters,
but rather emergent observables resulting from the collective behavior of the interacting active particles.
As such they depend on all system parameters, in particular $\phi$ and $\Pe$
(see also the discussion in Appendix~\ref{app:numerics:clusterfrac}).

\subsection{Numerical simulations}
\label{sec:results:numerics}

To test the relations~\eqref{eq:sigma=pa}, \eqref{eq:sigma=p} and~\eqref{eq:sigma=pa:MIPS:both},
we carry out Langevin simulations of the system~\eqref{eq:LE}, \eqref{eq:OUP}.
We here restrict ourselves to a brief overview of the methodology. Full details of the simulation parameters and methods are provided in Appendix~\ref{app:numerics}.

We adopt periodic boundary conditions
and choose an interaction potential [see below Eq.~\eqref{eq:force:def}] of the form
$U(r) = \frac{1}{5} [(\frac{2R}{r})^{32} - 1]^2 \, \Theta(2R - r)$.
The irreversibility $\sigma$ is calculated directly from the $\tau\to\infty$ limit of Eq.~\eqref{eq:sigma:1}
(cf.\ Eq.~\eqref{eq:sigma:simplified2}),
which was the starting point of our derivation.
The right-hand sides of Eqs.~\eqref{eq:sigma=pa}--\eqref{eq:sigma=pa:MIPS:both}, in turn,
are obtained from
the known (fixed) values of $N$, $T$, $\ta$, and $V_{\mathrm{p}}$
as well as
steady-state averages of the pressure~\eqref{eq:p}--\eqref{eq:pressures}.
The latter depend on the absolute particle positions,
but can be rewritten equivalently in terms of ``local'' quantities (forces, velocities, and particle distances),
cf.\ Eq.~\eqref{eq:pa:num} for $\pa$ and Eq.~\eqref{eq:pint:num} for $\pint$.
To address the artificial discontinuities in particle positions caused by the finite, periodic simulation box,
we compute the pressure numerically using those local formulations that remain continuous as particles exit
through one boundary and re-enter through the opposite side.
Likewise, the representation~\eqref{eq:sigma:simplified2} used for $\sigma$
only depends on such continuous quantities.
Furthermore, observing that the number of nearest neighbors for dense sphere packing is $6$ ($12$) in $d = 2$ ($d = 3$),
the value of $\phi_{\mathrm{d}}$ is calculated from~\eqref{eq:phi:dilute} by estimating
$n_{\mathrm{c}}$
as the fraction of particles that are surrounded by at least 5 (10) other particles within a distance of $R/10$;
we choose values slightly smaller than the maxima to account for occasional defects, see also the snapshots in Figs.~\ref{fig:PhasesPressIrrev2D}a and~\ref{fig:PhasesPressIrrev3D}a.

In $d=2$, the numerical results from Fig.~\ref{fig:PhasesPressIrrev2D}c demonstrate that the relation~\eqref{eq:sigma=pa} is satisfied excellently for small-to-moderate
$\phi \lesssim 0.2$ for all $\Pe$ (compare the orange and blue curves in all panels).
For larger values of $\Pe \gtrsim 30$, we even find almost perfect agreement up to the critical point, i.e., for $\phi \lesssim 0.35$.
Furthermore, relation~\eqref{eq:sigma=p} is confirmed equally well for sufficiently large $\Pe$ and small $\phi$ (compare the orange and black curves).
Finally, comparing the orange and purple lines, we observe that the refined relation~\eqref{eq:sigma=pa:MIPS} for the MIPS regime is satisfied
extremely well after the phase transition,
which manifests itself as a sudden drop in $\sigma$ (and $\pa$, $p$).
In fact, in $d = 2$ we find the relation~\eqref{eq:sigma=pa:MIPS}
to hold almost perfectly
for all $\phi$
if  $\Pe$ is sufficiently large
(third and fourth panels in Fig.~\ref{fig:PhasesPressIrrev2D}c).

In $d=3$, the numerical results from Fig.~\ref{fig:PhasesPressIrrev3D}c again
provide compelling evidence supporting the validity of Eq.~\eqref{eq:sigma=pa} 
for small $\phi \lesssim 0.15$ for all $\Pe$ (orange vs.\ blue curves),
and likewise for Eq.~\eqref{eq:sigma=p} (orange vs.\ black curves).
Similarly, the MIPS-corrected relation~\eqref{eq:sigma=pa:MIPS} is found to work
very well in the MIPS regime and at high
packing fractions (compare the orange and purple lines).
Furthermore, in both $d = 2$ and $d = 3$, we can confirm that the relation~\eqref{eq:sigma=pa:MIPS}
reproduces Eq.~\eqref{eq:sigma=pa} in the low-density limit as
the blue and purple (and orange) lines in Figs.~\ref{fig:PhasesPressIrrev2D}c and~\ref{fig:PhasesPressIrrev3D}c coincide for small $\phi$.

We observe that the
modifications due to the prefactor $\frac{\phi_{\mathrm{d}}}{\phi}$ in Eq.~\eqref{eq:sigma=pa:MIPS}
are significantly smaller in $d = 3$ than in $d = 2$ (compare the purple and blue lines in Figs.~\ref{fig:PhasesPressIrrev2D}c and~\ref{fig:PhasesPressIrrev3D}c).
In other words,
the correction factor $(1-n_{\mathrm{c}})/(1-n_{\mathrm{c}} \phi/\phi_{\mathrm{cp}})$
appears to be closer to unity in $d = 3$ than in $d = 2$.
We identify three factors contributing to this trend:
First, the closest packing fraction $\phi_{\mathrm{cp}}$ is smaller in $d = 3$,
implying that the ratio $\phi/\phi_{\mathrm{cp}}$
is closer to unity.
Second, the phase transition
occurs
at slightly larger $\phi$ in $d = 3$,
such that the smallest values of $\phi$ exhibiting MIPS,
for which the correction is the largest,
are closer to $\phi_{\mathrm{cp}}$ in $d = 3$ than in $d = 2$.
Third, at the onset of MIPS the fraction $n_{\mathrm{c}}$ of particles aggregated into clusters
is smaller in $d = 3$ than in $d = 2$
(see also Appendix~\ref{app:numerics:clusterfrac}).

Contrary to the situation in $d = 2$,
we find a range of intermediate packing fractions in $d = 3$, from $\phi \approx 0.2$ up to the
onset of MIPS,
for which none of our relations~\eqref{eq:sigma=pa}--\eqref{eq:sigma=pa:MIPS} is quantitatively accurate,
not even for large $\Pe$.
However, $\sigma$ and the right-hand sides of~\eqref{eq:sigma=pa} and~\eqref{eq:sigma=pa:MIPS} still exhibit the same qualitative behavior.
In this range of packing fractions collisions of three or more particles are likely to occur more frequently,
yet without forming large clusters,
so that such higher-order collisions
make a non-negligible contribution to $\sigma$.
We presume that these effects are stronger in $d=3$,
because there is generally more
space for higher-order collisions:
a particle can overlap simultaneously with up to 12 other particles in $d=3$,
whereas the maximum number of simultaneous collision partners is $6$ in $d = 2$.

\section{Discussion and conclusions}
\label{sec:discussion}

Our main results, relations~\eqref{eq:sigma=pa}--\eqref{eq:sigma=pa:MIPS:both},
provide a remarkable connection between the dynamical irreversibility of the steady state
of repulsively interacting active particles and the associated thermodynamic state variables, notably the pressure these particles exert on the container walls.
For a given system size $N$ and temperature $T$, it
shows that the
active pressure
can be used to infer
how (ir)reversible the spatial trajectories
of the active particle system are, i.e., 
how far the system appears to be away from equilibrium
\cite{Fodor:2016hff,Saw:2023cti}
when measuring the irreversibility of its dynamical history
\cite{Dabelow:2019iam, Dabelow:2021hia, Martin:2021sma}.

Irreversibility as defined in~\eqref{eq:Sigma} and~\eqref{eq:sigma:def} is
often interpreted as a non-equilibrium generalization of entropy production
\cite{Neri:2017sis, Fodor:2022ibe}.
However, \emph{a priori} this definition of $\sigma$ is unrelated to the usual thermodynamic definition of entropy
and should rather be seen as an information-theoretic (relative) entropy of steady-state path probabilities.
In contrast,
our results~\eqref{eq:sigma=pa}--\eqref{eq:sigma=pa:MIPS:both}
provide a proper thermodynamic interpretation
of irreversibility: The left-hand side
is the average irreversibility produced
in the steady state per unit time.
The right-hand sides of~\eqref{eq:sigma=pa}--\eqref{eq:sigma=pa:MIPS:both} express this ``non-equilibrium entropy production rate''
as a state function of the thermodynamic system parameters,
namely
the traditional state variables $N$ and $T$ as well as
the active pressure $\pa$.
In this spirit, $\sigma$ resembles a thermodynamic potential, but for a non-equilibrium steady state.
As such, the irreversibility is sensitive to phase changes in the system.
This becomes apparent in
the numerical results
in Figs.~\ref{fig:PhasesPressIrrev2D} and \ref{fig:PhasesPressIrrev3D},
showing that
$\sigma$ develops a discontinuity at the transition from the homogeneous to the MIPS-exhibiting coexistence phase.

Despite their similarity to equilibrium thermodynamic relations, the non-equilibrium character of our results~\eqref{eq:sigma=pa}--\eqref{eq:sigma=pa:MIPS:both}
becomes evident in a profound difference.
In equilibrium systems, entropy is a state function of the thermodynamic parameters, and can only change
when the system transitions from one thermodynamic state to another one by altering the state variables.
For non-quasistatic
transitions, such processes ``produce irreversibility''
during the transient relaxation to the new equilibrium state.
In contrast, an active matter system continuously dissipates energy. Its dynamical
evolution in the steady state will thus look more and more irreversible over time,
even in a pure trajectory-level description which does not resolve the self-propulsion mechanisms.
The relations~\eqref{eq:sigma=pa}--\eqref{eq:sigma=pa:MIPS:both} identify the
corresponding \emph{steady-state production rate} of irreversibility as a state function.
They reveal that variations of the thermodynamic parameters not only induce irreversibility production
during the transient relaxation to a new steady state,
but especially affect the rate at which irreversibility is produced in the latter.

In the form~\eqref{eq:sigma=p},
our findings provide
a direct and simple way to assess irreversibility experimentally at sufficiently small packing fractions and high activity
by measuring the total pressure $p$.
As demonstrated numerically (cf.\ especially Fig.~\ref{fig:PhasesPressIrrev2D}c),
this relation can hold up to the critical point and the onset of MIPS.

We derived the relations~\eqref{eq:sigma=pa}--\eqref{eq:sigma=pa:MIPS:both}
for the prototypical model of active Ornstein-Uhlenbeck particles (AOUPs),
exploiting the availability of analytical expressions for the irreversibility.
Nevertheless,
we expect that the essence of this link between the dynamics and thermodynamics of active matter holds much more generally
and especially
in models with a very similar phenomenology, such as ABPs and RTPs (cf.\ below Eq.~\eqref{eq:etaCorr}).
The reason is that
the derivation itself (cf.\ Appendices~\ref{app:rel1} and~\ref{app:rel2})
is largely based on kinematic arguments only.
None of these arguments depend on the explicit form of the interaction forces (see Eq.~\eqref{eq:force:def})
as long as they are repulsive and sufficiently stiff.
Similarly, most steps of the derivation do not rely on a specific model for the active fluctuations.
Indeed, in the derivation of relation~\eqref{eq:rel1} (Appendix~\ref{app:rel1}),
$\sqrt{2 \Da} \, \VEC\eta_i(t)$ in the Langevin equation~\eqref{eq:LE} could be replaced
by any other unbiased process
with a finite correlation time $\ta$ without affecting the main arguments.
The Gaussianity of $\VEC\eta_i(t)$ is exploited only when evaluating~\eqref{eq:rel2},
i.e., when reducing the correlations between the particle positions and fluctuating
forces at different time points to equal-time correlations.
We expect that relations analogous to Eq.~\eqref{eq:rel2} exist, at least approximately, for
other self-propulsion models, presumably with a modified numerical prefactor.
Likewise, our extension of the 
low-density
result~\eqref{eq:sigma=pa}
to the MIPS regime at high density and activity, Eq.~\eqref{eq:sigma=pa:MIPS:both},
is based on the observation that particles in large clusters are effectively arrested and thus cannot produce irreversibility.
This geometrical argument is model-independent, and
should again apply very generally to other active-matter models,
given the ubiquity of MIPS.

However, an explicit derivation for other models is currently not feasible
as Eq.~\eqref{eq:sigma:1}---the starting point of our analysis---has no known counterpart beyond the AOUP model.
Nevertheless, it
is plausible to expect some structural similarity to Eq.~\eqref{eq:sigma:1}, again at least approximately,
but with a model-specific memory kernel.
In particular,
the appearance of a scalar product between velocities $\dot{\VEC x}_i$ and forces $\VEC f_i$ is directly related to the form of the equation of motion~\eqref{eq:LE} and
how the relevant quantities transform under time reversal.
We also note that, even though choosing this starting point~\eqref{eq:sigma:1}
is inspired by stochastic thermodynamics and the questions raised
within this framework \cite{Jarzynski:2011eai, Seifert:2012stf},
our derivation does
not make use of its typical tools,
nor does it require the concepts of stochastic energetics \cite{Sekimoto:StochasticEnergetics}.
Rather, it essentially boils down to a careful
analysis of two-time correlations.

Ultimately, relations like~\eqref{eq:sigma=pa},
\eqref{eq:sigma=p}, and \eqref{eq:sigma=pa:MIPS:both}
are non-equilibrium
generalizations of the entropy as a thermodynamic state function.
In equilibrium, the entropy is at the foundations of thermodynamics
and the second law in particular.
As such, it is a crucial concept across the physical sciences and technical applications,
encoding the feasibility and efficiency of thermodynamic processes.
A
very important
class of such processes is
energy conversion,
during muscle contraction in biology,
during electrolysis in chemistry,
or in combustion engines in engineering, to name but a few examples.
Our novel interpretation of the trajectory-level irreversibility
may provide a similar basis for processes driven by or, more generally, involving active matter.
Notably, it may provide a first step towards studying thermodynamic processes with active matter
in parameter space (on a macroscopic level,
similarly to usual thermodynamic processes between equilibrium states),
rather than in trajectory space (microscopic level)
\cite{Pietzonka:2019aed,Ekeh:2020tcw,Fodor:2021aet,Datta:2022sla,Sorkin:2024slo}.
Provided that a suitable analog of the first law can be formulated as well,
this may ultimately allow
for optimizing such processes \cite{Gupta:2023ecp,Davis:2024amu},
for instance in active engines \cite{Krishnamurthy:2016msh, Pietzonka:2019aed, Datta:2022sla, GarciaMillan:2024ocl},
without the need to understand the microscopic details
of self-propulsion
and the generated trajectories.
To this end, it will be necessary to generalize our central insights to
situations in which externally controlled potentials or nonconservative forces are
present,
which can exchange
work with the active fluid.

\begin{acknowledgments}
R.E.\ acknowledges funding by the Swedish Research Council (Vetenskapsr{\aa}det)
under Grant No.~2020-05266 and Grant No.~638-2013-9243.
This research utilized Queen Mary's Apocrita HPC facility, supported by QMUL Research-IT \cite{King:2017ahp}.
\end{acknowledgments}

\appendix

\begin{widetext}

\section{Derivation of relation~(\ref{eq:sigma:2})}
\label{app:sigma}

Generalizing the formalism for a single
active Ornstein-Uhlenbeck particle (AOUP) from
Refs.~\cite{Dabelow:2019iam,Dabelow:2021iam}
to $N$ interacting AOUPs with trajectories running from initial time $\ti$ to final time $\tf$,
the log-ratio $\Sigma[\traj{\VEC{x}}]$ between forward and backward
probabilities, Eq.~\eqref{eq:Sigma}, can be calculated exactly as
\begin{subequations}
\label{eq:Irrev+Gamma}
\begin{equation}
\label{eq:Irrev}
\Sigma[\traj{\VEC{x}}] = \frac{1}{D} \sum_{i=1}^N
\int_{\ti}^{\tf} \!\! \d t \int_{\ti}^{\tf} \!\! \d t^\prime \;
	\frac{1}{\gamma} \dot{\VEC{x}}_i(t) \cdot \VEC{f}_i(\VEC{x}(t^\prime))
	\left[ \delta(t-t^\prime) - \frac{\Da}{D} \Gamma(t, t^\prime) \right]
	+ \Delta S_{\mathrm{sys}}[\traj{\VEC{x}}]
\, .
\end{equation}
Here, $\Delta S_{\mathrm{sys}}[\traj{\VEC{x}}]$ is the change in the system entropy between the initial and final system states,
and $\Gamma(t, t')$ is a memory kernel emerging from the correlated nature of the active fluctuations
(cf.~Eq.~\eqref{eq:etaCorr}).
If the initial state of the active fluctuations is chosen independently of the particle positions,
$\Gamma(t, t')$ is given by
\cite{Dabelow:2019iam,Dabelow:2021iam}
\begin{equation}
\label{eq:Gamma}
\Gamma(t, t') = \left(  \frac{\tau_*}{2 \ta^2} \right)
	\frac{ \kappa_+^2 e^{-| t - t' |/\tau_*} + \kappa_-^2 e^{-(2(\tf-\ti) - | t - t' |) / \tau_*}
		- \kappa_+ \kappa_- \left[ e^{-(t + t'-2\ti)/\tau_*} + e^{-(2\tf - t - t')/\tau_*} \right] }
	       { \kappa_+^2 - \kappa_-^2 e^{-2(\ti-\tf) / \tau_*} }
\, ,
\end{equation}
\end{subequations}
with $\tau_* = \ta / \sqrt{1 + \tfrac{\Da}{D}}$ as before and $\kappa_\pm := 1 \pm \sqrt{1 + \tfrac{\Da}{D}}$.
The expressions~\eqref{eq:Irrev+Gamma} provide an explicit and
exact form for the irreversibility of finite-time AOUP trajectories.

Our theoretical analysis is based 
on the average irreversibility production rate $\sigma$ in the steady state, as defined in~\eqref{eq:sigma:def}.
Note that we chose $-\tau_{\mathrm{i}}=\tau/2=\tau_{\mathrm{f}}$ for the initial and final time
for convenience, such that the steady-state limit $\tau_{\mathrm{i}} \to -\infty$
and $\tau_{\mathrm{f}} \to \infty$ is reached when performing $\tau \to \infty$.
The representation~\eqref{eq:sigma:1} of $\sigma$ in the main text follows straightforwardly from \eqref{eq:Irrev}
by observing that the system entropy $\Delta S_{\mathrm{sys}}$
is a time-intensive quantity, such that $\Delta S_{\mathrm{sys}}/\tau$ vanishes as $\tau \to \infty$.

To establish that $\sigma$ from~\eqref{eq:sigma:1} can be rewritten as in Eq.~\eqref{eq:sigma:2},
we first note that,
due to the conservative nature of the particle interactions $\VEC{f}_i$ [see below Eq.~\eqref{eq:force:def}],
the $\delta$-contribution
is a boundary term and thus vanishes in the steady state as well.
Moreover, as $\tau \to \infty$, the asymptotic form of the memory
kernel is $\Gamma(t, t') \sim \tau_* / (2 \ta^2) \, e^{-| t - t' | / \tau_*}$
(by letting $\ti \to -\infty$ and $\tf \to +\infty$ in~\eqref{eq:Gamma}).
Hence, Eq.~\eqref{eq:sigma:1} simplifies to
\begin{equation}
\label{eq:sigma:simplified}
\sigma = -\frac{\Da}{2 D^2 \ta^2} \sum_{i=1}^N \lim_{\tau\to\infty} \frac{\tau_*}{\tau}
	\int_{-\tau/2}^{\tau/2} \!\! \d t \int_{-\tau/2}^{\tau/2} \!\! \d t^\prime \; e^{-| t - t' | / \tau_*}
	\frac{1}{\gamma} \left\langle \dot{\VEC{x}}_i(t) \cdot \VEC{f}_i(\VEC{x}(t^\prime)) \right\rangle
\, .
\end{equation}
We now split the domain of the $t$ integral
into $[-\tau/2, t']$ and $[t', \tau/2]$,
\begin{equation}
\sigma = -\frac{\Da}{2 D^2 \ta^2} \sum_{i=1}^N \lim_{\tau\to\infty} \frac{\tau_*}{\tau}
	\int_{-\tau/2}^{\tau/2} \!\! \d t' \left\langle \frac{1}{\gamma}  \VEC{f}_i(\VEC{x}(t')) \cdot
		\left[ \int_{-\tau/2}^{t'} \d t \, e^{-(t'- t) /\tau_*} \dot{\VEC{x}}_i(t)
		     + \int_{t'}^{\tau/2} \d t \, e^{-(t - t') / \tau_*} \dot{\VEC{x}}_i(t) \right]
	\right\rangle
\, .
\end{equation}
Next, we substitute $s = t' - t$ in the first integral and $s = t - t'$ in the second one,
\begin{equation}
\sigma = -\frac{\Da}{2 D^2 \ta^2 \gamma} \sum_{i=1}^N \lim_{\tau\to\infty} \frac{\tau_*}{\tau}
	\int_{-\tau/2}^{\tau/2} \!\! \d t' \left[ \int_{0}^{\tau/2+t'}\!\!\!\!\!\!\!\!\! \d s \, e^{-s /\tau_*} \left\langle  \VEC{f}_i(\VEC{x}(t')) \cdot
		 \dot{\VEC{x}}_i(t'-s) \right\rangle
		     + \int_{0}^{\tau/2-t'} \!\!\!\!\!\!\!\!\! \d s \, e^{-s / \tau_*} \left\langle  \VEC{f}_i(\VEC{x}(t')) \cdot \dot{\VEC{x}}_i(t'+s) \right\rangle \right]
\, .
\end{equation}
The correlation functions in the integrands are independent of $t'$ because the steady-state average only depends on the difference between the two time points,
such that we can replace $t'$ by an arbitrary reference time $t_0$.
Changing the order of the $t'$ and $s$ intgrals, we then obtain
\begin{equation}
\begin{aligned}
\sigma &= -\frac{\Da}{2 D^2 \ta^2 \gamma} \sum_{i=1}^N \lim_{\tau\to\infty} \frac{\tau_*}{\tau}
	\left[ \int_{0}^{\tau}\! \d s \, e^{-s /\tau_*} \left\langle  \VEC{f}_i(\VEC{x}(t_0)) \cdot
		 \dot{\VEC{x}}_i(t_0-s) \right\rangle \int_{-\tau/2 + s}^{\tau/2} \!\!\!\!\!\! \d t'
		 \right. \\ & \qquad\qquad\qquad\qquad\qquad\qquad \left.
		     + \int_{0}^{\tau} \! \d s \, e^{-s / \tau_*} \left\langle  \VEC{f}_i(\VEC{x}(t_0)) \cdot \dot{\VEC{x}}_i(t_0+s) \right\rangle \int_{-\tau/2}^{\tau/2 - s} \!\!\!\!\!\! \d t' \right]
\, .
\end{aligned}
\end{equation}
Both $t'$ integrals yield a factor of $\tau - s$.
Evaluating the $\tau\to\infty$ limit, we thus arrive at
\begin{equation}
\label{eq:sigma:simplified2}
\sigma = -\frac{\Da \tau_*}{2 D^2 \ta^2 \gamma} \sum_{i=1}^N \int_{0}^{\infty}\! \d s \, e^{-s /\tau_*}
	\left[ \left\langle  \VEC{f}_i(\VEC{x}(t_0)) \cdot
		 \dot{\VEC{x}}_i(t_0-s) \right\rangle 
		     + \left\langle  \VEC{f}_i(\VEC{x}(t_0)) \cdot \dot{\VEC{x}}_i(t_0+s) \right\rangle \right]
\, .
\end{equation}
Finally, we integrate by parts in $s$ in both integrals,
which leads to
\begin{equation}
\begin{aligned}
	\sigma &= \frac{\Da \tau_*}{2 D^2 \ta^2 \gamma} \sum_{i=1}^N
		\left\{ \vphantom{\int_0^\infty}
		\left. e^{-s/\tau_*} \left[ \langle \bm f_i(\bm x(t_0)) \cdot \bm x_i(t_0 - s) \rangle - \langle \bm f_i(\bm x(t_0)) \cdot \bm x_i(t_0 + s) \rangle \right] \right|_{s=0}^\infty
		\right. \\ & \qquad\qquad\qquad\qquad \left.				
		 + \frac{1}{\tau_*} \int_0^\infty \! \d s \, e^{-s/\tau_*} \left[ \langle \bm f_i(\bm x(t_0)) \cdot \bm x_i(t_0 - s) \rangle - \langle \bm f_i(\bm x(t_0)) \cdot \bm x_i(t_0 + s) \rangle \right] \right\}
\end{aligned}
\end{equation}
The boundary terms vanish and we are left with
\begin{equation}
\sigma =  \frac{\Da}{2 D^2 \ta^2 \gamma} \sum_{i=1}^N 
	 \int_0^\infty \d s \, e^{-s / \tau_*}
	 \left[  
	 	\langle \VEC{x}_i(t_0-s) \cdot  \VEC{f}_i(\VEC{x}(t_0)) \rangle
	       -\langle \VEC{x}_i(t_0+s) \cdot \VEC{f}_i(\VEC{x}(t_0)) \rangle
	\right] ,
\end{equation}
which is equivalent to Eq.~\eqref{eq:sigma:2} after a shift of the reference time $t_0$.

\section{Derivation of relation~(\ref{eq:rel1})}
\label{app:rel1}

We provide the analytical arguments to corroborate that relation \eqref{eq:rel1} is valid for small
packing fractions $\phi$,
i.e., we analyze the two-time correlations
$\sum_{i=1}^N \langle \VEC{x}_i(t_0) \cdot \VEC{f}_i(\VEC{x}(t_0 \pm s)) \rangle$.
We start by
decomposing the total interaction
force $\VEC{f}_i(\VEC{x})$ acting on particle $i$ into its pairwise interactions
$\VEC{f}(\VEC{x}_i-\VEC{x}_j)$
with all other particles $j$
according to Eq.~\eqref{eq:force:def}, i.e.,
\begin{subequations}
\begin{align}
\label{eq:force:def:app}
\VEC{f}_i(\VEC{x})  &= \sum_{j (\neq i)} \VEC{f}(\VEC{x}_i-\VEC{x}_j)
\, ,
\\
\label{eq:force:rev}
\VEC{f}(\VEC{x}_i-\VEC{x}_j)  &= -\VEC{f}(\VEC{x}_j-\VEC{x}_i)
\, .
\end{align}
\end{subequations}
It follows that
\begin{equation}
\label{eq:step1}
\frac{1}{\gamma} \sum_{i=1}^N \langle \VEC{x}_i(t_0) \cdot \VEC{f}_i(\VEC{x}(t_0 \pm s)) \rangle
= \frac{1}{\gamma} \sum_{i<j}  \langle \VEC{r}_{ij}(t_0) \cdot \VEC{f}(\VEC{r}_{ij}(t_0 \pm s)) \rangle
\, ,
\end{equation}
where we introduced the
distance vectors
$\VEC{r}_{ij} := \VEC{x}_i - \VEC{x}_j$
between particles $i$ and $j$.
The equations of motion for the $\VEC{r}_{ij}$ are obtained from \eqref{eq:LE} via
$\dot{\VEC{r}}_{ij}(t) = \dot{\VEC{x}}_i(t)-\dot{\VEC{x}}_j(t)$,
\begin{equation}
\label{eq:LEr:exact}
	\dot{\VEC{r}}_{ij}(t) = \frac{1}{\gamma} \left[ \VEC{f}_i(\VEC{x}(t)) - \VEC{f}_j(\VEC{x}(t)) \right] + \VEC{\alpha}_{ij}(t)
\end{equation}
with the combined thermal and active fluctuations
\begin{equation}
\VEC{\alpha}_{ij}(t) := \VEC{\alpha}_i(t) - \VEC{\alpha}_j(t)
\, , \quad
\VEC{\alpha}_i(t) := \sqrt{2D} \, \VEC{\xi}_{i}(t) + \sqrt{2\Da} \, \VEC{\eta}_{i}(t)
\, .
\end{equation}
As explained in the main text,
simultaneous collisions of more than two particles become exceedingly rare for small packing fractions $\phi$.
For any fixed particle $i$,
we therefore assume that at most one other particle $j$ contributes in the sum in~\eqref{eq:force:def:app} at any given time.
We emphasize that this still admits collisions of different pairs $(i, j)$ at \emph{different} times.
For small $\phi$, we can thus replace $\VEC{f}_i(\VEC{x}(t)) - \VEC{f}_j(\VEC{x}(t))$ in~\eqref{eq:LEr:exact} by $2 \VEC{f}(\VEC{r}_{ij}(t))$, utilizing~\eqref{eq:force:rev} as well.
Therefore, we obtain
\begin{equation}
\label{eq:LEr}
\dot{\VEC{r}}_{ij}(t)
= \frac{2}{\gamma} \VEC{f}(\VEC{r}_{ij}(t)) + \VEC{\alpha}_{ij}(t)
\, .
\end{equation}

The particle-particle interaction force
$\VEC{f}(\VEC{r}_{ij})$ is always directed along the
distance vector $\VEC{r}_{ij}$,
(cf.~the definition of $\VEC{f}(\VEC{r})$ in the main text below Eq.~\eqref{eq:force:def}), i.e.,
\begin{equation}
\label{eq:f||e}
\VEC{f}(\VEC{r}_{ij}) \parallel \VEC{r}_{ij}
\, .
\end{equation}
Introducing the unit vector $\VEC{e}_{\VEC{r}_{ij}} := \VEC{r}_{ij} / \lvert \VEC{r}_{ij} \rvert$
pointing in the collision direction, we
can thus split the equation of motion \eqref{eq:LEr} into a component parallel
to $\VEC{e}_{\VEC{r}_{ij}}$ and a component perpendicular to it:
\begin{subequations}
\label{eq:LEr:decomposition}
\begin{align}
\label{eq:LEr:parallel}
\dot{\VEC{r}}_{ij}^\parallel(t) &= \frac{2}{\gamma} \VEC{f}(\VEC{r}_{ij}(t)) + \VEC{\alpha}_{ij}^\parallel(t)
\, , \\
\label{eq:LEr:perp}
\dot{\VEC{r}}_{ij}^\perp(t) &= \VEC{\alpha}_{ij}^\perp(t)
\, .
\end{align}
\end{subequations}
Here, vectors parallel (perpendicular) to
$\VEC{e}_{\VEC{r}_{ij}}$ are indicated by the superscript $\parallel$  ($\perp$),
i.e. $\VEC{\alpha}_{ij}^\parallel(t) := [\VEC{\alpha}_{ij}(t) \cdot \VEC{e}_{\VEC{r}_{ij}(t)} ] \VEC{e}_{\VEC{r}_{ij}(t)}$
and $\VEC{\alpha}_{ij}^\perp(t) := \VEC{\alpha}_{ij}(t) - \VEC{\alpha}_{ij}^\parallel(t)$, etc.
Note that the decomposition~\eqref{eq:LEr:decomposition}
is time dependent and stochastic because $\VEC{e}_{\VEC{r}_{ij}(t)}$ is a stochastic process.

As mentioned in the main text
below Eq.~\eqref{eq:force:def}, we consider short-ranged, strongly repulsive
particle-particle interaction forces
with $\VEC{f}(\VEC{r}_{ij}) \neq 0$ if $|\VEC{r}_{ij}| \leq 2R$ and $\VEC{f}(\VEC{r}_{ij}) = 0$ if $|\VEC{r}_{ij}| > 2R$.
This hard-core-like character of
$\VEC{f}(\VEC{r}_{ij})$ implies
that, in very good approximation,
$\dot{\VEC{r}}_{ij}^\parallel(t) \approx 0$
whenever the particles $i$ and $j$ collide,
i.e., whenever $\lvert \VEC{r}_{ij} \rvert \leq 2R$.
(Note that there is no restriction on the perpendicular components $\VEC{r}_{ij}^\perp$.)
We can thus rewrite \eqref{eq:LEr:parallel} as
\begin{equation}
\label{eq:forceValpha}
\frac{1}{\gamma} \VEC{f}(\VEC{r}_{ij}(t)) \approx -\frac{1}{2} \VEC{\alpha}_{ij}^\parallel(t) \,
\chi(\lvert \VEC{r}_{ij}(t) \rvert)
\end{equation}
with the characteristic function
\begin{equation}
\chi(r) := \left\{
	\begin{array}{cl}
		1 & \mbox{if } r \leq 2R \\
		0 & \mbox{otherwise}
	\end{array}
\right.
\, .
\end{equation}
Substituting~\eqref{eq:forceValpha} into the right-hand side of~\eqref{eq:step1},
the correlation function can be approximated by
\begin{equation}
\label{eq:step2}
\frac{1}{\gamma} \sum_{i=1}^N \langle \VEC{x}_i(t_0) \cdot \VEC{f}_i(\VEC{x}(t_0 \pm s)) \rangle
= -\frac{1}{2}
	\sum_{i<j} \langle \VEC{r}_{ij}(t_0) \cdot \VEC{\alpha}_{ij}^\parallel(t_0 \pm s) \, 
\chi(\lvert \VEC{r}_{ij}(t_0 \pm s) \rvert ) 
 \rangle \,.
\end{equation}

Recalling the definition of the projection onto $\VEC{e}_{\VEC{r}_{ij}(t)}$ (see below~\eqref{eq:LEr:perp})
and defining $r_{ij}(t) := \lvert \VEC{r}_{ij}(t) \rvert$ for short,
the steady-state average on the right-hand side can be written as
\begin{equation}
\label{eq:prefactor:start}
\langle \VEC{r}_{ij}(t_0) \cdot \VEC{\alpha}_{ij}^\parallel(t_0 \pm s) \, \chi(r_{ij}(t_0 \pm s)) 
 \rangle
=\langle [\VEC{r}_{ij}(t_0) \cdot \VEC{e}_{\VEC{r}_{ij}(t_0 \pm s)}]
[\VEC{\alpha}_{ij}(t_0 \pm s) \cdot  \VEC{e}_{\VEC{r}_{ij}(t_0 \pm s)}] \, \chi(r_{ij}(t_0 \pm s))
 \rangle \,.
\end{equation}

Next we carry out
the average over the orientation vector $\VEC{e}_{\VEC{r}_{ij}(t_0 \pm s)}$ of the
particle distance at time $t_0 \pm s$.
For our spherical active particles with radially symmetric interactions obeying~\eqref{eq:f||e},
the steady-state distribution
does not depend on a specific (reference) direction in space.
Moreover,
as long as the particles move around independently, i.e., without colliding,
their distribution will be independent of the relative orientation $\VEC{e}_{\VEC{r}_{ij}}$.
However, due to the additional factor $\chi(r_{ij}(t_0 \pm s))$
enforcing
$r_{ij} \leq 2R$,
the average in~\eqref{eq:prefactor:start} is restricted to ``collision configurations'' of the two particles.
Since we are focusing on the regime
dominated by pair collisions,
the particles can only collide if they move towards each other,
i.e., if $\VEC{\alpha}_{ij}(t_0 \pm s) \cdot \VEC{e}_{\VEC{r}_{ij}(t_0 \pm s)} < 0$.
As soon as they move apart, they will not overlap anymore and
$r_{ij} > 2R$ (see also above Eq.~\eqref{eq:forceValpha}).
To lowest order in $\phi$, we can then combine the aforementioned homogeneity of the
independent particle movement \emph{before} the collision
with the condition $\VEC{\alpha}_{ij}(t_0 \pm s) \cdot \VEC{e}_{\VEC{r}_{ij}(t_0 \pm s)} < 0$
required to \emph{initiate} and maintain a collision.
We thus average uniformly over $\VEC{e}_{\VEC{r}_{ij}(t_0 \pm s)}$ on a half sphere
and find that
\begin{equation}
\label{eq:prefactor:step0}
\langle [\VEC{r}_{ij}(t_0) \cdot \VEC{e}_{\VEC{r}_{ij}(t_0 \pm s)}]
	   [\VEC{\alpha}_{ij}(t_0 \pm s) \cdot  \VEC{e}_{\VEC{r}_{ij}(t_0 \pm s)}] \, \chi(r_{ij}(t_0 \pm s)) \rangle
= \frac{1}{2d} \langle \VEC{r}_{ij}(t_0) \cdot \VEC{\alpha}_{ij}(t_0 \pm s) \, \chi(r_{ij}(t_0 \pm s)) \rangle \,.
\end{equation}
Here we exploited the relation
\begin{equation}
\label{eq:UnitSphereAvg}
\int_{\lvert \VEC{n} \rvert = 1} \frac{\d^d n}{S_d} \, (\VEC{a}\cdot\VEC{n})(\VEC{b}\cdot\VEC{n}) = \frac{\VEC{a}\cdot\VEC{b}}{d}
\,
\end{equation}
for integrals over the unit sphere
in $d$ dimensions,
where $S_d$ is its surface area, and
$\VEC{a}$ and $\VEC{b}$ are arbitrary vectors \cite{Kania:2015spf}.
The additional factor of $\frac{1}{2}$ on the right-hand side of~\eqref{eq:prefactor:step0} results from the condition $\VEC{\alpha}_{ij}(t_0 \pm s) \cdot \VEC{e}_{\VEC{r}_{ij}(t_0 \pm s)} < 0$, which restricts the domain to a half sphere.
The approximation leading to relation \eqref{eq:prefactor:step0} amounts to neglecting correlations
between $\VEC{e}_{\VEC{r}_{ij}}$ and $\VEC{\alpha}_{ij}$
that build up in the collision region, and
thus introduces a relativ error of $\mathcal{O}(V_{\mathrm{p}}/V)$
(with the particle volume $V_{\mathrm{p}}$).
Substituting~\eqref{eq:prefactor:step0} into~\eqref{eq:prefactor:start},
we arrive at
\begin{equation}
\label{eq:prefactor:step1}
\langle \VEC{r}_{ij}(t_0) \cdot \VEC{\alpha}_{ij}^\parallel(t_0 \pm s) \, \chi(r_{ij}(t_0 \pm s)) \rangle
= \frac{1}{2d} \langle \VEC{r}_{ij}(t_0) \cdot \VEC{\alpha}_{ij}(t_0 \pm s) \, \chi(r_{ij}(t_0 \pm s)) \rangle
\left[ 1 + \mathcal{O}\!\left(\!\tfrac{V_{\mathrm{p}}}{V} \!\right) \right]
\end{equation}

On the right-hand side, we now switch back to the original single-particle variables, such that
\begin{align}
\label{eq:prefactor:step2}
& \langle \VEC{r}_{ij}(t_0) \cdot \VEC{\alpha}_{ij}^\parallel(t_0 \pm s) \, \chi(r_{ij}(t_0 \pm s) \rangle
\nonumber \\ \qquad
& = \frac{1}{2d}
\langle [\VEC{x}_{i}(t_0)-\VEC{x}_{j}(t_0)]
   \cdot [\VEC{\alpha}_{i}(t_0 \pm s) - \VEC{\alpha}_{j}(t_0 \pm s)] \, \chi(|\VEC{x}_{i}(t_0 \pm s) - \VEC{x}_{j}(t_0 \pm s)|) \rangle
\nonumber \\ \qquad
&
= \frac{1}{2d} \left\langle
 \left[
	   \VEC{x}_{i}(t_0) \cdot \VEC{\alpha}_{i}(t_0 \pm s) 
	+ \VEC{x}_{j}(t_0) \cdot \VEC{\alpha}_{j}(t_0 \pm s) 
-  \VEC{x}_{i}(t_0) \cdot \VEC{\alpha}_{j}(t_0 \pm s)
	- \VEC{x}_{j}(t_0) \cdot \VEC{\alpha}_{i}(t_0 \pm s)
\right]
	\chi(|\VEC{x}_{i}(t_0 \pm s) - \VEC{x}_{j}(t_0 \pm s)|)
	\right\rangle
\nonumber \\ \qquad
& =  \frac{1}{d}  \langle \VEC{x}_{i}(t_0) \cdot \VEC{\alpha}_{i}(t_0 \pm s) \, \chi(|\VEC{x}_{i}(t_0 \pm s) - \VEC{x}_{j}(t_0 \pm s)|)
 \rangle
      - \frac{1}{d} \langle \VEC{x}_{i}(t_0) \cdot \VEC{\alpha}_{j}(t_0 \pm s) \, \chi(|\VEC{x}_{i}(t_0 \pm s) - \VEC{x}_{j}(t_0 \pm s)|)
\rangle
\, 
\end{align}
up to a relative error of $\mathcal{O}(\frac{V_{\mathrm{p}}}{V})$.
Note that in the last step we exploited that the steady-state probability as well as
the expression $\chi(|\VEC{x}_{i}(t_0 \pm s) - \VEC{x}_{j}(t_0 \pm s)|)$
are invariant under exchange of the particle indices $i$ and $j$.

The first average in \eqref{eq:prefactor:step2} is now over the steady-state probability density
$p(\VEC{x}_{i}',t_0;\VEC{x}_{i},\VEC{\alpha}_{i},\VEC{x}_j,t_0 \pm s)$
for finding particle $i$ at position $\VEC{x}_{i}'$ at time $t_0$ and at position $\VEC{x}_i$
with fluctuating velocity $\VEC{\alpha}_i$ at time $t_0 \pm s$,
while also having a different particle $j$ at position $\VEC{x}_j$ at the same time $t_0 \pm s$.
We isolate the average over the particle position $\VEC{x}_j$ at time $t_0 \pm s$ and write
\begin{multline}
\label{eq:prefactor:step3}
\frac{1}{d} \langle \VEC{x}_{i}(t_0) \cdot \VEC{\alpha}_{i}(t_0 \pm s) \, \chi(|\VEC{x}_{i}(t_0 \pm s) - \VEC{x}_{j}(t_0 \pm s)|) \rangle
= \frac{1}{d} \int \d^d x_i' \, \d^d x_i \, \d^d \alpha_i \; p(\VEC{x}_{i}',t_0;\VEC{x}_{i},\VEC{\alpha}_{i},t_0 \pm s) \,
	 \VEC{x}_{i}' \cdot \VEC{\alpha}_i
\\
\mbox{} \times
    \int \d^d x_j \, p(\VEC{x}_j,t_0 \pm s | \VEC{x}_{i}',t_0;\VEC{x}_{i},\VEC{\alpha}_{i},t_0 \pm s ) \,
    	\chi(|\VEC{x}_{i} - \VEC{x}_{j}|)
\, .
\end{multline}
For small packing fractions $\phi$, the conditional density
$p(\VEC{x}_j,t_0 \pm s | \VEC{x}_{i}',t_0;\VEC{x}_{i},\VEC{\alpha}_{i},t_0 \pm s )$ for
the spatial position of particle $j$ is uniform and independent of the state of particle $i$
almost everywhere in the accessible volume, except in
the vicinity of the
region $|\VEC{x}_{i}-\VEC{x}_j| \lesssim 2R$, where the particles interact and
the density depends on all quantities explicitly.
Approximating
$p(\VEC{x}_j,t_0 \pm s | \VEC{x}_{i}',t_0;\VEC{x}_{i},\VEC{\alpha}_{i},t_0 \pm s )$
by a uniform density 
thus introduces an error of the order $V_{\mathrm{p}}/V$,
\begin{equation}
\label{eq:papprox}
p(\VEC{x}_j,t_0 \pm s | \VEC{x}_{i}',t_0;\VEC{x}_{i},\VEC{\alpha}_{i},t_0 \pm s ) 
= \frac{1}{V} [ 1 + \mathcal{O}(V_{\mathrm{p}}/V) ]
\, ,
\end{equation}
because the two densities differ only around positions close to $\VEC{x}_i'$ and $\VEC{x}_i$ within volumes
of the order of the particle volume, but are identical everywhere else.
This is the second key consequence of our assumption of small packing fractions $\phi$.
Adopting~\eqref{eq:papprox} in \eqref{eq:prefactor:step3}, we can perform the integration over $\VEC{x}_j$
and obtain
\begin{equation}
\label{eq:prefactor:step3.5}
\frac{1}{d} \langle \VEC{x}_{i}(t_0) \cdot \VEC{\alpha}_{i}(t_0 \pm s) \,
\chi(|\VEC{x}_{i}(t_0 \pm s) - \VEC{x}_{j}(t_0 \pm s)|)
 \rangle
= \frac{2^d V_{\mathrm{p}}}{d V}
\left[ 1 \!+\! \mathcal{O}\!\left(\!\tfrac{V_{\mathrm{p}}}{V} \!\right) \right]
\int \d^d x_i' \, \d^d x_i \, \d^d \alpha_i \; p(\VEC{x}_{i}',t_0;\VEC{x}_{i},\VEC{\alpha}_{i},t_0 \pm s) \,
	\VEC{x}_{i}' \cdot \VEC{\alpha}_i
\, .
\end{equation}
The prefactor $2^d V_{\mathrm{p}}$ corresponds to the volume designated by the characteristic function $\chi(r)$,
i.e., $\pi (2R)^2=2^2 V_{\mathrm{p}}$ in $d=2$ and $(4\pi/3)(2R)^3=2^3 V_{\mathrm{p}}$ in $d=3$.
The remaining integral over $\VEC{x}_i$ in~\eqref{eq:prefactor:step3.5} amounts to a trivial factor of $1$ due to normalization,
leaving us with
\begin{subequations}
\label{eq:prefactor:step4}
\begin{equation}
\frac{1}{d} \langle \VEC{x}_{i}(t_0) \cdot \VEC{\alpha}_{i}(t_0 \pm s) \, \chi(|\VEC{x}_{i}(t_0 \pm s) - \VEC{x}_{j}(t_0 \pm s)|) \rangle
=  \frac{2^d V_{\mathrm{p}}}{dV} \langle \VEC{x}_{i}(t_0) \cdot \VEC{\alpha}_{i}(t_0 \pm s) \rangle
\left[ 1 + \mathcal{O}\!\left(\!\tfrac{V_{\mathrm{p}}}{V} \!\right) \right]
\,.
\end{equation}
An analogous argument leads to a similar relation for the second term in \eqref{eq:prefactor:step2},
\begin{equation}
\frac{1}{d} \langle \VEC{x}_{i}(t_0) \cdot \VEC{\alpha}_{j}(t_0 \pm s) \, \chi(|\VEC{x}_{i}(t_0 \pm s) - \VEC{x}_{j}(t_0 \pm s)|) \rangle
=  \frac{2^d V_{\mathrm{p}}}{dV} \langle \VEC{x}_{i}(t_0) \cdot \VEC{\alpha}_{j}(t_0 \pm s) \rangle
\left[ 1 + \mathcal{O}\!\left(\!\tfrac{V_{\mathrm{p}}}{V} \!\right) \right]
\,.
\end{equation}
\end{subequations}

Using~\eqref{eq:prefactor:step2} and~\eqref{eq:prefactor:step4},
we return to Eq.~\eqref{eq:step2} and obtain
\begin{align}
&
\frac{1}{\gamma} \sum_{i=1}^N \langle \VEC{x}_i(t_0) \cdot \VEC{f}_i(\VEC{x}(t_0 \pm s)) \rangle
= -\frac{2^{d-1} V_{\mathrm{p}}}{dV} \sum_{i<j}
	\left[
		\langle \VEC{x}_{i}(t_0) \cdot \VEC{\alpha}_{i}(t_0 \pm s) \rangle
	      - \langle \VEC{x}_{i}(t_0) \cdot \VEC{\alpha}_{j}(t_0 \pm s) \rangle
	\right]
	\left[ 1 + \mathcal{O}\!\left(\!\tfrac{V_{\mathrm{p}}}{V} \!\right) \right]
\nonumber \\ \qquad
& \quad
 = -\frac{2^{d-2} (N-1) V_{\mathrm{p}}}{dV} 
 \left[ 1 + \mathcal{O}\!\left(\!\tfrac{(N-1) V_{\mathrm{p}}}{V} \!\right) \right]
 \sum_{i} \langle \VEC{x}_{i}(t_0) \cdot \VEC{\alpha}_{i}(t_0 \pm s) \rangle
+ \frac{2^{d-2} V_{\mathrm{p}}}{dV} \sum_{i \neq j} \langle \VEC{x}_{i}(t_0) \cdot \VEC{\alpha}_{j}(t_0 \pm s) \rangle 
	\left[ 1 + \mathcal{O}\!\left(\!\tfrac{V_{\mathrm{p}}}{V} \!\right) \right]
\,.
\end{align}
The expressions $(N-1)V_{\mathrm{p}}/V \approx NV_{\mathrm{p}}/V$ in the first term can be identified with
the packing fraction $\phi$.
The absolute error of the first term is thus $\mathcal{O}(\phi^2)$.
Similarly as above,
the correlation function $\langle \VEC{x}_i(t_0) \cdot \VEC{\alpha}_j(t_0 \pm s) \rangle$ for \emph{different} particles in the second term is of order $\mathcal{O}(\frac{V_{\mathrm{p}}}{V})$.
Summing over all $N(N-1)$ contributions thus leads to a term of order $\mathcal{O}(\frac{N^2 V_{\mathrm{p}}}{V})$.
Together with the prefactor $\frac{V_{\mathrm{p}}}{V}$,
the second term is thus of order $\mathcal{O}(\phi^2)$.
We therefore conclude that
\begin{equation}
\frac{1}{\gamma} \sum_{i=1}^N \langle \VEC{x}_i(t_0) \cdot \VEC{f}_i(\VEC{x}(t_0 \pm s)) \rangle
= -\frac{2^{d-2}}{d} \phi \sum_{i} \langle \VEC{x}_{i}(t_0) \cdot \VEC{\alpha}_{i}(t_0 \pm s) \rangle \, \left(1 + \mathcal{O}(\phi) \right)
\, ,
\end{equation}
which is relation \eqref{eq:rel1} in the main text.

\section{Derivation of relation~(\ref{eq:rel2})}
\label{app:rel2}
The derivation of relation \eqref{eq:rel2} is based on
\emph{Novikov's theorem}
\cite{Novikov:1965frf}.
Given an unbiased Gaussian stochastic process $\alpha(t)$ with
correlation function $C(t, s) := \langle \alpha(t)\alpha(s) \rangle$ and an arbitrary functional
$\mathcal{F}[\alpha](t_0)$ depending on this stochastic process up to time $t_0$, the Novikov theorem states that
\begin{align}
\label{eq:Novikov}
\langle \mathcal{F}[\alpha](t_0) \alpha(t) \rangle
& = \int_{-\infty}^{t_0} \d s \, C(t,s) \left\langle \frac{\delta \mathcal{F}[\alpha](t_0)}{\delta \alpha(s)} \right\rangle
\, .
\end{align}

To show Eq.~\eqref{eq:rel2},
we choose $\alpha(t)$ as one of the spatial components $\nu$ of the combined thermal and active fluctuations
of an arbitrary particle $i$,
\begin{subequations}
\begin{equation}
\label{eq:alpha}
\alpha(t) = \sqrt{2D} \, \xi_i^\nu(t) + \sqrt{2\Da} \, \eta_i^\nu(t)
\, ,
\end{equation}
such that
\begin{equation}
\label{eq:C}
C(t,s) = 2D \, \delta(t-s) + \frac{\Da}{\ta} e^{-|t-s|/\ta}
\, ,
\end{equation}
\end{subequations}
independently of $\nu$ and $i$ (cf.\ Eq.~\eqref{eq:etaCorr}).
The functional  $\mathcal{F}[\alpha](t_0)$, in turn, is the $\nu$th component of the $i$th particle's position
trajectory in the stationary state, from initial time $t=-\infty$ to final time $t=t_0$,
i.e., $\mathcal{F}[\alpha](t_0) = x_i^\nu(t_0)$.
For the following proof we keep general $\mathcal{F}[\alpha](t_0)$ though, and only set $t_0=0$ for convenience.

We thus want to calculate the integrals
\begin{align}
\label{eq:Ipm}
I_\pm :=
\int_0^\infty \d s \, \langle \mathcal{F}[\alpha](0) \, \alpha(\pm s) \rangle \, e^{-s/\tau_*}
= \int_0^\infty \d s \, e^{-s/\tau_*} \int_{-\infty}^0 \d s'  \, C(\pm s,s')  
	\left\langle \frac{\delta \mathcal{F}[\alpha](0)}{\delta \alpha(s')}  \right\rangle
\, .
\end{align}
To proceed, we plug in the expression \eqref{eq:C} for the correlations and consider
the two signs separately.

For the plus sign,
we thus find
\begin{equation}
	I_+	= \int_0^\infty \d s \, e^{-s/\tau_*} \int_{-\infty}^0 \d s'  \left[ 2 D \, \delta(s - s') + \frac{\Da}{\ta} \, e^{-\lvert s - s' \rvert/\ta} \right]  
	\left\langle \frac{\delta \mathcal{F}[\alpha](0)}{\delta \alpha(s')}  \right\rangle .
\end{equation}
The term involving the $\delta$ function does not contribute since $s > s'$,
and the remaining integrals factorize.
Evaluating the $s$ integral,
we therefore obtain
\begin{align}
\label{eq:+s:s'}
I_+
&= \frac{\Da \tau_*}{\ta + \tau_*}
	\int_{-\infty}^0 \d s' \, e^{s'/\ta}
	\left\langle \frac{\delta \mathcal{F}[\alpha](0)}{\delta \alpha(s')}  \right\rangle
\, .
\end{align}
Exploiting first the definition~\eqref{eq:alpha} and then Novikov's theorem~\eqref{eq:Novikov},
we can furthermore rewrite
\begin{equation}
\label{eq:NovikovInverse}
\int_{-\infty}^0 \d s' \, e^{s'/\ta}
\left\langle \frac{\delta \mathcal{F}[\alpha](0)}{\delta \alpha(s')}  \right\rangle
= \int_{-\infty}^0 \d s' \, e^{s'/\ta}
\left\langle \frac{\delta \mathcal{F}[\alpha](0)}{\delta \sqrt{2\Da}\eta_i^\nu(s')}  \right\rangle
= \frac{\ta}{\Da} \langle \mathcal{F}[\alpha](0) \sqrt{2\Da}\eta_i^\nu(0) \rangle \,.
\end{equation}
Adopting this result in~\eqref{eq:+s:s'} and recalling $\tau_* = \ta / \sqrt{1 + \Da/D}$ (see below Eq.~\eqref{eq:sigma:2}),
we find
\begin{equation}
\label{eq:+s}
	I_+ = \frac{\ta}{\sqrt{1 + \Da/D} + 1} \langle \mathcal{F}[\alpha](0) \sqrt{2\Da}\eta_i^\nu(0) \rangle \,.
\end{equation}

Turning to the minus sign in~\eqref{eq:Ipm},
we have
\begin{equation}
	I_- = \int_0^\infty \d s \, e^{-s/\tau_*} \int_{-\infty}^0 \d s'  \left[ 2 D \, \delta(s + s') + \frac{\Da}{\ta} \, e^{-\lvert s + s' \rvert/\ta} \right]  
	\left\langle \frac{\delta \mathcal{F}[\alpha](0)}{\delta \alpha(s')}  \right\rangle 
\end{equation}
after substituting~\eqref{eq:C}.
For the contribution involving the $\delta$ function,
we can directly evaluate the $s$ integral.
For the remaining term, we split the domain of the $s$ integral at $-s'$,
such that
\begin{equation}
	I_- = \int_{-\infty}^0 \d s' \left\langle \frac{\delta \mathcal{F}[\alpha](0)}{\delta \alpha(s')} \right\rangle \left[ 2D \, e^{s'/\tau_*}  
+   \frac{\Da}{\ta} e^{s'/\ta}
	    \int_0^{-s'} \d s \, e^{s(1/\ta-1/\tau_*)}  
	+  \frac{\Da}{\ta} e^{-s'/\ta} \int_{-s'}^\infty \d s \, e^{-s(1/\ta + 1/\tau_*)} 
\right] .
\end{equation}
Evaluating the $s$ integrals,
we obtain
\begin{equation}
\label{eq:I-:step}
	I_- = \int_{-\infty}^0 \d s' \left\langle \frac{\delta \mathcal{F}[\alpha](0)}{\delta \alpha(s')} \right\rangle \left[ 2D \, e^{s'/\tau_*}  
+   \frac{\Da \tau_*}{\tau_* - \ta} \left( e^{s'/\tau_*} - e^{s'/\ta} \right)  
	+  \frac{\Da \tau_*}{\tau_* + \ta} e^{s'/\tau_*} \right] .
\end{equation}
With $\tau_* = \ta / \sqrt{1 + \Da/D}$,
we observe that $\frac{\tau_*}{\tau_*-\ta} + \frac{\tau_*}{\tau_*+\ta} = -\frac{2D}{\Da}$,
such that the terms proportional to $e^{s'/\tau_*}$ in~\eqref{eq:I-:step} cancel each other exactly.
Hence we are left with
\begin{equation}
	I_- = \frac{\Da \tau_*}{\ta - \tau_*}
	\int_{-\infty}^0 \d s' \, e^{s'/\ta}
	\left\langle \frac{\delta \mathcal{F}[\alpha](0)}{\delta \alpha(s')}  \right\rangle .
\end{equation}
Using~\eqref{eq:NovikovInverse} and $\tau_* = \ta / \sqrt{1 + \Da/D}$ once again,
we find
\begin{equation}
\label{eq:-s}
	I_- = \frac{\ta}{\sqrt{1 + \Da/D} - 1} \langle \mathcal{F}[\alpha](0) \sqrt{2\Da}\eta_i^\nu(0) \rangle \,.
\end{equation}
Combining Eqs.~\eqref{eq:+s} and~\eqref{eq:-s} and switching back to an arbitrary reference time $t_0$,
we finally obtain
\begin{equation}
	I_\pm = \frac{\ta}{\sqrt{1 + \Da/D} \pm 1} \langle \mathcal{F}[\alpha](t_0) \sqrt{2\Da}\eta_i^\nu(t_0) \rangle \,.
\end{equation}
Recalling the definition~\eqref{eq:Ipm}, substituting $\mathcal{F}[\alpha](t_0) = x^\nu_i(t_0)$ and summing over $\nu = 1, \ldots, d$,
this proves the relation~\eqref{eq:rel2}.

\end{widetext}

\section{Derivation of relation~(\ref{eq:sigma=p})}
\label{app:pa}

To complete the arguments used in the main text when deriving Eq.~\eqref{eq:sigma=p},
we have to show that $\frac{\pa}{\po} = \mathcal{O}(\Pe^2)$ and $\frac{\pa}{\pint} = \mathcal{O}(\phi^{-1})$.

To establish $\frac{\pa}{\po} = \mathcal{O}(\Pe^2)$,
it is convenient to rewrite the active pressure
as defined in~\eqref{eq:pa} in the form
\begin{equation}
\label{eq:pa:num}
	\pa = \frac{N \gamma \Da}{V} + \frac{\ta}{d V} \sum_i \langle \bm f_i(\bm x) \cdot \sqrt{2 \Da} \, \bm\eta_i \rangle 
\, .
\end{equation}
The proof that Eqs.~\eqref{eq:pa} and \eqref{eq:pa:num} are equivalent will be given below.
Dividing by $\po=N\gamma D/V$ (see Eq.~\eqref{eq:p0}) on both sides,
we obtain
\begin{equation}
\label{eq:pa/po}
	\frac{\pa}{\po}
		= \Pe^2  \left[ 1 + \frac{\sqrt{2\Da}}{\Da} \frac{\ta}{N \gamma d}
\sum_i \langle \VEC{f}_i(\VEC{x}(t_0)) \cdot \VEC{\eta}_i(t_0) \rangle \right]
\, .
\end{equation}
Since we are interested in the limit of small packing fractions,
we can adopt the line of reasoning from Appendix~\ref{app:rel1}
and effectively replace
$\VEC{f}_i(\VEC{x}(t_0))/\gamma$ by $-\phi \VEC{\alpha}_i(t_0) = -\phi [\sqrt{2D} \, \VEC{\xi}_i (t_0)- \sqrt{2 \Da} \, \VEC{\eta}_i(t_0) ] = -\phi \sqrt{2 \Da} [\VEC{\eta}_i(t_0) + \mathcal{O}(\Pe^{-1})]$
up to a constant prefactor.
For $\Pe \gg 1$, the
leading contribution of
the pressure ratio $\frac{\pa}{\po}$, Eq.~\eqref{eq:pa/po},
is thus indeed $\mathcal{O}(\Pe^2)$.

To estimate the ratio $\frac{\pint}{\pa}$,
we consider the representations~\eqref{eq:pint} and~\eqref{eq:pa}.
Adopting the same argument as above, we can replace
$\VEC{f}_i(\VEC{x}(t_0))/\gamma$ by $-\phi \sqrt{2 \Da} [\VEC{\eta}_i(t_0) + \mathcal{O}(\Pe^{-1})]$
in~\eqref{eq:pint} up to a constant prefactor.
Comparing to~\eqref{eq:pa}, we conclude that $\frac{\pint}{\pa} = \mathcal{O}(\phi)$ to leading order if $\phi \ll 1$ and $\Pe \gg 1$.

We now demonstrate how to derive the representation
\eqref{eq:pa:num} of the active pressure from its definition \eqref{eq:pa}.
Similar calculations for ABPs and RTPs
can be found in Refs.~\cite{Solon:2015ppe,Solon:2015pin,Winkler:2015vps}.
Let us
consider some arbitrary, finite reference time $t_0$ in the steady-state average in~\eqref{eq:pa}.
Substituting $\VEC{x}_i(t_0) = \VEC{x}_i(-\infty) + \int_{-\infty}^{t_0} \d t \, \dot{\VEC{x}}_i(t)$
and exploiting that $\langle \VEC{\eta}_i(t_0) \cdot \VEC{x}_i(-\infty) \rangle = 0$,
we obtain
\begin{equation}
\label{eq:pa:num:start}
	\pa = \frac{\gamma \sqrt{2 \Da}}{d V} \sum_i \int_{-\infty}^{t_0} \!\! \d t \, 
		\langle \VEC{\eta}_i(t_0) \cdot \dot{\VEC{x}}_i(t) \rangle
\, .
\end{equation}
Next we use the Langevin equation~\eqref{eq:LE} to replace $\dot{\VEC{x}}_i(t)$.
Observing that $\langle \VEC{\eta}_i(t_0) \cdot \bm\xi_i(t) \rangle = 0$ by definition,
we find
\begin{equation}
\label{eq:ActivePressure:EOMRepl}
\begin{aligned}
	\pa &= \frac{2 \Da \gamma}{d V} \sum_i \int_{-\infty}^{t_0} \!\! \d t \, \langle \VEC{\eta}_i(t_0) \cdot \VEC{\eta}_i(t) \rangle
	\\ & \quad
		+ \frac{\sqrt{2 \Da}}{d V} \sum_i \int_{-\infty}^{t_0} \!\! \d t \,\langle \VEC{\eta}_i(t_0) \cdot \VEC{f}_i(\VEC{x}(t)) \rangle .
\end{aligned}
\end{equation}

In the first term of this relation,
we can directly calculate the integral using~\eqref{eq:etaCorr}.
The result is
\begin{equation}
\label{eq:ActivePressure:EOMRepl:1}
	\frac{2 \Da \gamma}{d V} \sum_i \int_{-\infty}^{t_0} \!\! \d t \langle \VEC{\eta}_i(t_0) \cdot \VEC{\eta}_i(t) \rangle
		= \frac{N \gamma \Da}{V}
\, .
\end{equation}
For the second term in~\eqref{eq:ActivePressure:EOMRepl},
we note that we can express the active fluctuations as
\begin{equation}
\label{eq:eta:stationary}
	\VEC{\eta}_i(t_0) = \frac{1}{\ta} \int_{-\infty}^{t_0} \d s \, e^{-(t_0-s) / \ta} \, \VEC{\zeta}_i(s)
\, ,
\end{equation}
which is the formal steady-state integral representation of the Ornstein-Uhlenbeck process~\eqref{eq:OUP}.
We can thus rewrite
\begin{equation}
\begin{aligned}
	 &\int_{-\infty}^{t_0} \! \d t \, \langle \VEC{\eta}_i(t_0) \cdot \VEC{f}_i(\VEC{x}(t)) \rangle
	\\	&\;
	= \frac{1}{\ta} \int_{-\infty}^{t_0} \! \d t \, e^{-(t_0 - t)/\ta} \int_{-\infty}^{t_0} \! \d s \, e^{-(t-s)/\ta} \langle \VEC{\zeta}_i(s) \cdot \VEC{f}_i(\VEC{x}(t)) \rangle
\, .
\end{aligned}
\end{equation}
Since  $\langle \VEC{\zeta}_i(s) \cdot \VEC{f}_i(\VEC{x}(t)) \rangle = 0$ for $s > t$, the
domain of the $s$ integral is effectively restricted to an upper bound  $t$ instead of $t_0$.
Hence we can use~\eqref{eq:eta:stationary} once again
and find
\begin{equation}
\begin{aligned}
	&\int_{-\infty}^{t_0} \!\! \d t \, \langle \VEC{\eta}_i(t_0) \cdot \VEC{f}_i(\VEC{x}(t)) \rangle
	\\ & \quad
		= \int_{-\infty}^{t_0} \!\! \d t \, e^{-(t_0 - t) / \ta} \langle \VEC{\eta}_i(t) \cdot \VEC{f}_i(\VEC{x}(t)) \rangle .
\end{aligned}
\end{equation}
Because of stationarity, the average on the right-hand side does not actually 
depend on the integration variable $t$, such that
\begin{equation}
\label{eq:ActivePressure:EOMRepl:2}
	\int_{-\infty}^{t_0} \!\! \d t \, \langle \VEC{\eta}_i(t_0) \cdot \VEC{f}_i(\VEC{x}(t)) \rangle
		= \ta \langle \VEC{\eta}_i(t_0) \cdot \VEC{f}_i(\VEC{x}(t_0)) \rangle .
\end{equation}
Substituting~\eqref{eq:ActivePressure:EOMRepl:1} and~\eqref{eq:ActivePressure:EOMRepl:2}
in~\eqref{eq:ActivePressure:EOMRepl},
we finally recover Eq.~\eqref{eq:pa:num}.

\section{Numerical simulations}
\label{app:numerics}

We carry out Langevin simulations of the dynamical system~\eqref{eq:LE}--\eqref{eq:OUP}
in two and three dimensions, using a standard Euler algorithm.

\subsection{System specifications}

The active particles are confined in a
hypercubic
box of volume $V = L^d$ with periodic boundary conditions.
The hard-core two-body interactions as defined in and below~Eq.~\eqref{eq:force:def}
are modeled by a steep power-law potential with a finite-range cutoff of the form
\cite{Stenhammar:2014pba,Solon:2015ppe, Solon:2015pin, Levis:2017abe, Digregorio:2018fpd, Martin:2021sma}
\begin{equation}
\label{eq:IntPotentialNum}
	U(r) = k \left[ \left( \tfrac{2R}{r} \right)^{32} - 1 \right]^{2} \Theta(2R - r) \,,
\end{equation}
where $\Theta(x)$ denotes the Heaviside step function.
We fix the parameter $k=0.2$ and the particle radius $R = 1$.
For the thermal environment we choose temperature $T = 0.01$ and dynamic viscosity $\mu = 0.1$, leading to a Stokes friction coefficient $\gamma = 6\pi\mu R \approx 1.88$.
The active correlation time is $\ta = 4 R^2 / (3D) \approx 251$ (corresponding to the inverse rotational diffusion coefficient of a sphere of radius $R$ \cite{Bechinger:2016api}).
Variations of the packing fraction $\phi$ (cf.\ above Eq.~\eqref{eq:LE}) and P\'{e}clet number $\Pe$ (cf.\ above Eq.~\eqref{eq:OUP}) are achieved by adjusting the linear box dimension $L$ and the active diffusivity $\Da$, respectively.

All numerical results are obtained
from systems of $N = 5000$ particles in $d = 2$ and $N = 20\,000$ particles in $d=3$.
The integration time steps $\Delta t$ are chosen relatively small due to the steep interaction potential~\eqref{eq:IntPotentialNum}.
We adopt $\Delta t \leq 2 \times 10^{-4}$,
guaranteeing that the maximal displacement $\lvert \bm x_i(t + \Delta t) - \bm x_i(t) \rvert$
remains below $R/50 = 0.02$.
The data points in Figs.~\ref{fig:PhasesPressIrrev2D}c and~\ref{fig:PhasesPressIrrev3D}c are averages over six independent repetitions of the simulations.
The error bars visualize the corresponding standard error.

\subsection{Initial state}
\label{app:numerics:init}

The initial particle positions are generated as follows:
For small packing fractions $\phi$,
the particles are scattered uniformly across the available volume
such that no two particles overlap,
i.e., discarding any candidate positions that would lead to an overlap.
For larger $\phi$, this approach becomes infeasible.
Instead, we then divide the space into $N$ or more equally sized unit cells
(hexagonal in 2D, face-centered cubic in 3D),
assign a cell to each particle,
and sample their positions randomly within their respective cells.

The active fluctuations $\bm\eta_i$ are initialized independently of the particle positions by sampling from
their marginal steady-state distribution,
\begin{equation}
	p_{\mathrm{ss}}(\bm\eta_i) = \sqrt{ \tfrac{\ta}{\pi} } e^{-\ta \bm\eta_i^2}
\, ,
\end{equation}
obtained from Eq.~\eqref{eq:OUP} by recognizing that it describes diffusive motion in a
``quadratic potential'' $\VEC{\eta}_i^2/(2\ta)$ at ``thermal energy'' $1/(2\ta^2)$.

Starting from these initial conditions,
we thermalize the system by running the simulations up to $t = 10\,000$.
Then we take the dynamics from $t = 10\,000 \ldots 12\,000$ as our steady-state sample
with trajectories of duration $\tau = 2000$.

\subsection{Irreversibility}
\label{app:numerics:sigma}

To estimate the irreversibility~\eqref{eq:sigma:def},
we numerically evaluate
$\sigma$ as given in Eq.~\eqref{eq:sigma:simplified2}.
Since this quantity is non-local in time,
we need to store trajectory segments around the current simulation time $t_0$
to evaluate $\sigma$.
Observing the exponential suppression of large time separations $s$,
we restrict the domain of the $s$ integrals in~\eqref{eq:sigma:simplified2}
and only integrate $s \in [0, -\tau_* \ln \epsilon_*]$,
where $\epsilon_* = 10^{-6}$ is the target relative precision.
In other words,
we ignore trajectory segments where
$e^{-s/\tau_*} \lesssim \epsilon_*$.
In particular, we thus only need to store histories of duration $-2 \tau_* \, \ln(\epsilon_*)$.
Furthermore,
we exploit the smoothening effect of $e^{-s / \tau_*}$ in~\eqref{eq:sigma:simplified2}
and allow for a larger time step $\Delta t_* = n_* \Delta t$ for the $s$ integrals.
We choose $n_*$ between $5$ and $40$,
with
larger values for larger $\tau_*$.

\begin{figure}
\includegraphics[scale=1]{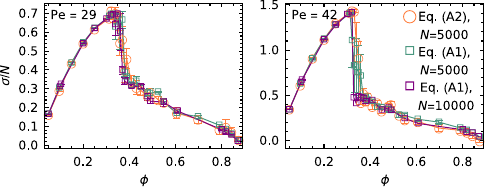}
\caption{Cross-check of numerical methods to calculate the irreversibility $\sigma$.
Plots show the irreversibility per particle, $\sigma / N$, as a function of the packing fraction $\phi$ for $\Pe = 29$ (left) and $\Pe = 42$ (right) in $d=2$).
Irreversibility is calculated via Eq.~\eqref{eq:sigma:simplified2} for $N = 5000$ particles
(orange; reference method used in the main text)
or via Eq.~\eqref{eq:Irrev+Gamma} for $N = 5000$ (green) and $N = 10\,000$ (purple);
see Appendix~\ref{app:numerics:sigma} for details.}
\label{fig:IrrevCmp}
\end{figure}

We also cross-checked this approach for selected configurations in $d = 2$ with a direct evaluation of the full finite-time trajectory expression $\Sigma - \Delta S_{\mathrm{sys}}$ as
given in Eq.~\eqref{eq:Irrev+Gamma}.
We recall that this expression is the exact path-probability log-ratio for a setup where $\bm\eta$ is initialized in the stationary marginal distribution, independent of the particle positions $\bm x$ \cite{Dabelow:2019iam},
which is precisely the initialization chosen in the simulations (cf.\ Appendix~\ref{app:numerics:init}).
Hence we calculate the double integral $\Sigma(\tau)$ for trajectories of durations $\tau = 10\,000, 10\,500, 11\,000, 11\,500, 12\,000$ (i.e., including the thermalization phase)
and estimate $\sigma$ by fitting the linear model $\Sigma(\tau) = \Sigma_0 + \sigma \tau$ to the so-obtained data points.

The two approaches are compared in Fig.~\ref{fig:IrrevCmp}:
The orange data points correspond to estimates via Eq.~\eqref{eq:sigma:simplified} for $N = 5000$ particles,
which is the same method as used in Fig.~\ref{fig:PhasesPressIrrev2D}.
The green data points were obtained via Eq.~\eqref{eq:Irrev+Gamma} as outlined in the previous paragraph and for $N = 5000$.
To test for potential finite-size effects,
we also include estimates obtained via Eq.~\eqref{eq:Irrev+Gamma} for $N = 10\,000$ particles.
Within statistical uncertainties,
these three independent estimates agree.

\subsection{Pressure}

The definitions~\eqref{eq:pint} and~\eqref{eq:pa} of $\pint$ and $\pa$ are numerically inconvenient in combination with the periodic boundary conditions
because the quantities exhibit discontinuities when a particle ``leaves'' the volume on one side of the box and ``re-enters'' on the other side,
which would need to be corrected.
However, both expressions can be rewritten in terms of purely local quantities,
which are continuous across the boundary and thus numerically more convenient to deal with.

For the active pressure~\eqref{eq:pa}, such a local representation was derived in Appendix~\ref{app:pa}, namely Eq.~\eqref{eq:pa:num}.

For the interaction pressure~\eqref{eq:pint},
we can substitute the definition~\eqref{eq:force:def} of the interaction forces
and exploit that $\bm f(\bm x_i - \bm x_j) = -\bm f(\bm x_j - \bm x_i)$.
This leads to
\begin{equation}
\label{eq:pint:num}
	\pint = \frac{1}{d V} \sum_{i < j} \langle \bm f(\bm x_i - \bm x_j) \cdot (\bm x_i - \bm x_j) \rangle \,,
\end{equation}
where $\bm x_i - \bm x_j$ is interpreted as the ``local'' distance vector,
i.e., the shortest connection between all periodic copies of both particles.
Note that this is the only configuration that can possibly contribute because of the short-range nature of $\bm f(\bm x_i - \bm x_j)$.

\subsection{MIPS cluster size}
\label{app:numerics:clusterfrac}

\begin{figure}
\centering
\includegraphics[scale=1]{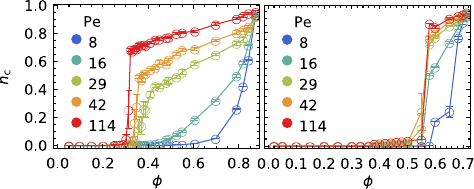}
\caption{Fraction of clustering particles $n_{\mathrm{c}}$ as a function of the packing fraction for various P\'{e}clet numbers in $d = 2$ (left) and $3$ (right),
see Appendix~\ref{app:numerics:clusterfrac} for details.
Error bars indicate one standard deviation.}
\label{fig:clusterfrac}
\end{figure}

\begin{figure}
\centering
\includegraphics[scale=1]{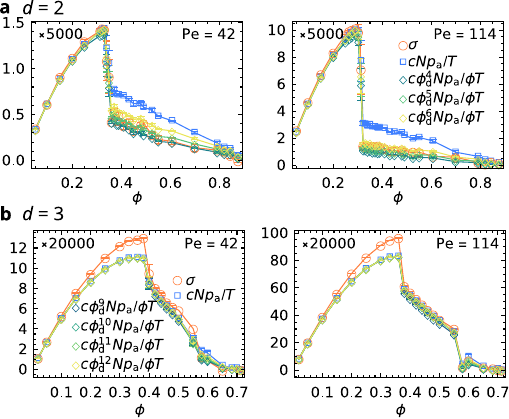}
\caption{Influence of the cluster threshold parameter $k_*$ on the relation~\eqref{eq:sigma=pa:MIPS} between
irreversibility and the thermodynamic state variables in (a) $d = 2$ and (b) $d = 3$.
The plots show the irreversibility $\sigma$ (orange), the right-hand side of Eq.~\eqref{eq:sigma=pa} (blue),
and the right-hand side of Eq.~\eqref{eq:sigma=pa:MIPS} for $\phi_{\mathrm{d}} = \phi_{\mathrm{d}}^{k_*}$
with various values of $k_*$ used
in Eq.~\eqref{eq:Nc} to calculate $\phi_{\mathrm{d}}$ according to Eq.~\eqref{eq:phi:dilute}.
All other parameters as in Figs.~\ref{fig:PhasesPressIrrev2D} and~\ref{fig:PhasesPressIrrev3D}.}
\label{fig:PhasesPressIrrev:ClusterfracComparison}
\end{figure}

To determine the dilute packing fraction $\phi_{\mathrm{d}}$ in the MIPS regime as defined in Eq.~\eqref{eq:phi:dilute},
we need to estimate the number $N_{\mathrm{c}}$ and fraction $n_{\mathrm{c}} = N_{\mathrm{c}}/N$ of particles in large clusters.
We define $N_{\mathrm{c}}$ as the number of particles that have a least $k_*$ neighbors within a distance of $R/10$,
\begin{equation}
\label{eq:Nc}
	N_{\mathrm{c}} = \left\lvert \left\{ i : \sum_{j(\neq i)} \Theta\!\left(2R + \frac{R}{10} - \lvert \bm x_i - \bm x_j \rvert\right) \geq k_* \right\} \right\rvert
\, ,
\end{equation}
where $|\{\cdot\}|$ denotes the cardinality of the set $\{\cdot\}$.
For given values of the system parameters,
we then calculate and average Eq.~\eqref{eq:Nc} over the simulated steady-state trajectories
using thresholds of $k_* = 5$ in $d = 2$ and $k_* = 10$ in $d = 3$.
The thresholds are chosen slightly below the coordination numbers for dense packing ($6$ and $12$, respectively) such that $N_{\mathrm{c}}$ comprises essentially all arrested particles in the bulk of large clusters,
which may sometimes exhibit small defects (see, for example, Figs.~\ref{fig:PhasesPressIrrev2D}a and~\ref{fig:PhasesPressIrrev3D}a).
The resulting fractions of clustering particles $n_{\mathrm{c}}$ for various $\phi$ and $\Pe$ are shown in Fig.~\ref{fig:clusterfrac}.

We observe that in $d = 3$ (right panel of Fig.~\ref{fig:clusterfrac}), the
clustering fractions $n_{\mathrm{c}}$ stay close to zero even after the onset of MIPS
at $\phi \approx 0.35$ (see right panel of Fig.~\ref{fig:alpha} or Fig.~\ref{fig:PhasesPressIrrev3D}b)
until they jump to high values around $\phi \approx 0.55$.
This indicates that the clusters in $d = 3$ 
just after the transition to the MIPS phase
are not densely packed.
We cannot currently rule out that this might be a finite-size artifact, but combined
with the results from Fig.~\ref{fig:PhasesPressIrrev3D}c an alternative explanation is warranted.
Interestingly, in the rightmost panel of Fig.~\ref{fig:PhasesPressIrrev3D}c)
we observe a second drop in the pressure (black curve) at around the same packing fraction
$\phi \approx 0.55$ at which the clustering fraction $n_{\mathrm{c}}$ jumps from close to zero to large values.
This might indicate that at $\phi \approx 0.55$
there is another ``phase transition'' or crossover (within the MIPS phase)
related to a re-structuring or ``freezing'' of the particle arrangement
within the clusters \cite{Turci:2021psa}.
We point out that the quantity $n_{\mathrm{c}}$ is
not designed to capture such (or similar) ``phase transitions'',
it merely measures
the fraction
of particles which are completely immobilized by a large number of nearest neighbors.
The physical intuition behind this definition (explained in detail in Sec.~\ref{sec:results:MIPS})
is strongly supported by the fact that the drop in the dilute packing fraction $\phi_{\mathrm{d}}$,
corresponding to the sudden increase of $n_{\mathrm{c}}$ at $\phi \approx 0.55$, nicely correlates with the
drop in the irreversibility production rate at about the same value of the packing fraction
(see the purple curve in the rightmost panel of Fig.~\ref{fig:PhasesPressIrrev3D}c).

Since there is some freedom in choosing $k_*$,
we also checked the influence of various reasonable choices on the validity of our relationship between the irreversibility and thermodynamic state variables in the MIPS regime, Eq.~\eqref{eq:sigma=pa:MIPS:both}.
To this end, we denote by $\phi_{\mathrm{d}}^{k_*}$ the estimate of $\phi_{\mathrm{d}}$ obtained from Eq.~\eqref{eq:phi:dilute} by using $n_{\mathrm{c}} = N_{\mathrm{c}}/N$ from~\eqref{eq:Nc} with the specified value of $k_*$.
Fig.~\ref{fig:PhasesPressIrrev:ClusterfracComparison} shows
the left- and right-hand sides of Eq.~\eqref{eq:sigma=pa:MIPS} (along with the uncorrected relation~\eqref{eq:sigma=pa}) for
choices of $k_*$ between $4$ and $6$ in $d = 2$ and between $9$ and $12$ in $d = 3$.
The resulting differences are small,
such that the spurious dependence on $k_*$ is negligible.

\subsection{Phase diagram}
\label{app:numerics:alpha}

\begin{figure}
\centering
\includegraphics[scale=1]{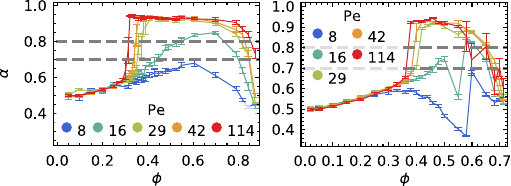}
\caption{Giant number fluctuations exponent $\alpha$ as a function of the packing fraction for various P\'{e}clet numbers in $d = 2$ (left) and $3$ (right),
see Appendix~\ref{app:numerics:alpha} for details.
Error bars indicate one standard deviation.}
\label{fig:alpha}
\end{figure}

The phase diagrams in Figs.~\ref{fig:PhasesPressIrrev2D}b and~\ref{fig:PhasesPressIrrev3D}b
are determined based on the so-called giant number fluctuations exponent $\alpha$
\cite{Fily:2012aps,Palacci:2013lcl}
as the order parameter. It is defined as follows:
Consider a random (hypercubic) sub-volume $v$ of the total box of volume $V$.
We denote by $\bar N_v$ and $\Delta N_v$ the mean and standard deviation, respectively, of the number of particles in such a randomly chosen sub-volume $v$.
Since the choice is random,
the mean number is $\bar N_v = N v / V$.
The standard deviation, however, varies depending on whether or not the particles cluster in certain regions of the space.
The scaling exponent $\alpha$ of $\Delta N_v$ with $\bar N_v$,
\begin{equation}
\label{eq:NumFlucScaling}
	\Delta N_v \propto (\bar N_v)^\alpha \,,
\end{equation}
determines the inhomogeneity of the particle distribution:
For a spatially homogeneous distribution,
we have $\Delta N_v \propto \sqrt{\bar N_v}$ ($\alpha = \frac{1}{2})$.
If the particles accumulate in densly packed clusters instead,
we find $\Delta N_v \propto \bar N_v$ ($\alpha = 1$).

Numerically, we determine $\alpha$ from snapshots of the simulated steady-state trajectories
at time intervals of $\tau / 200 = 10$.
For each snapshot and each target mean number $\bar N_v$,
we sample one random subspace of volume $v = \bar N_v V / N$,
and count the actual number of particles $N_v$.
We estimate $\Delta N_v$ by calculating the empirical standard deviation of those $N_v$ over all snapshots.
We then extract $\alpha$ as the slope of a linear least-squares fit to the $(\log \bar N_v, \log \Delta N_v)$ pairs for $\bar N_v$ in the range of 5 to 500.
We classify the homogeneous phase by $\alpha < 0.7$ and the coexistence phase by $\alpha > 0.8$.
The resulting values underlying the phase diagrams in Figs.~\ref{fig:PhasesPressIrrev2D}b and~\ref{fig:PhasesPressIrrev3D}c
are displayed in Fig.~\ref{fig:alpha}.

\end{document}